\documentclass[preprint,showpacs,preprintnumbers,amsmath,amssymb]{revtex4-1}
\usepackage{graphicx, latexsym, amssymb, amsmath, color, multirow, mathrsfs, booktabs, ifpdf}
\usepackage{dcolumn}
\usepackage{bm}

\def\al{$\alpha$}
\def\aN{$\alpha$-nucleus\ }
\def\aa{$\alpha + \alpha$\ }

\def\AA{nucleus-nucleus\ }

\begin{document}
\title{Folding model study of the elastic \aa scattering at low energies}
\author{Ngo Hai Tan}
\author{Nguyen Hoang Phuc}
\author{Dao T. Khoa}
\affiliation{\centerline{Institute for Nuclear Science \& Technology, Vinatom} 
 \centerline{179 Hoang Quoc Viet Road, Cau Giay, Hanoi.} }
\date{\today}
\begin{abstract}
The folding model analysis of the elastic \aa scattering at the incident energies 
below the reaction threshold of 34.7 MeV (in the lab system) has been done using 
the well-tested density dependent versions of the M3Y interaction and realistic 
choices for the $^4$He density. Because the absorption is negligible at the energies 
below the reaction threshold, we were able to probe the \aa optical potential at 
low energies quite unambiguously and found that the \aa overlap density used to
construct the density dependence of the M3Y interaction is strongly distorted by the 
Pauli blocking. This result gives possible explanation of a long-standing 
inconsistency of the double-folding model in its study of the elastic \aa and \aN 
scattering at low energies using the same realistic density dependent M3Y interaction. 
\end{abstract}
\maketitle

\section{Introduction}
\label{intro} The knowledge about the \aa interaction at low energies is of fundamental
importance due, in part, to diversities of the \al -cluster phenomena in nuclear physics, 
where one has to deal with configurations of two or more \al-particles interacting 
with each other. Thanks to the robust, tightly bound structure of the (spin- and isospin
zero) $^4$He nucleus, the elastic \aa scattering cross section has been measured 
quite accurately during the sixties and seventies of the last century, and numerous 
phase-shift analyses were made based on these cross sections \cite{Ali,Afzal}. The key 
quantity in an optical model (OM) study of elastic \aa scattering is the \aa optical 
potential (OP) that has been treated either phenomenologically \cite{Ali} or evaluated
microscopically from the (two-body) nucleon-nucleon (NN) interaction between nucleons 
bound in the two interacting \al-particles \cite{Buck,SatLove}. In terms of the quantum 
mechanical treatment of elastic \aa scattering, the resonating group method (RGM) 
(see, e.g., Refs.~\cite{Buck, Ho91}) is the most rigorous approach that takes into 
account the full antisymmetrization of the total wave function of the scattering system. 
The one-body wave equation for the relative wave function $\chi(R)$ is then constructed 
with a nonlocal RGM potential kernel. An accurate localization approximation 
has also been developed \cite{Ho91} to yield a local optical potential $U(R)$ to be used 
in the standard OM equation to determine $\chi(R)$. Given a complicated treatment of the 
nonlocal exchange kernel, only a \emph{density independent} NN interaction in the 
Gaussian form could be used as the effective interaction in the RGM calculation. 
The Pauli blocking effects have also been studied rigorously in the fish-bone model 
for the \aa potential \cite{Day11}. In a somewhat less rigorous way, the double-folding 
model (see, e.g., Refs.~\cite{SatLove,Kho00,Kho95,Kho97,Kho07}) 
determines the OP for the \aa system as 
\begin{equation}
 U=\sum_{i \in \alpha_1;j\in \alpha_2}[\langle ij|v_{\rm D}|ij\rangle 
 + \langle ij |v_{\rm EX}|ji\rangle],
\label{e1}
\end{equation}           
where $v_{\rm D(EX)}$ are the direct and exchange parts of the effective 
NN interaction between nucleons in the first \al-particle and those in the 
second one. The antisymmetrization gives rise to the exchange term in Eq.~(\ref{e1}) 
that is, in general, nonlocal in the coordinate space. To have a local double-folded 
OP, an accurate local approximation for the exchange potential has been developed 
\cite{Kho00,Kho97,Kho07}, which allowed the use of some realistic density dependent
NN interaction. Among different choices of the effective NN interaction, the original
density independent M3Y interactions \cite{Be77,An83} have been used with some 
success in the double-folding calculations of the heavy-ion (HI) optical potential at low 
energies \cite{SatLove}, where the data are sensitive only to the potential at the surface 
because of the strong absorption. However, in cases of refractive (rainbow) \AA 
scattering where the elastic data are sensitive to the OP over a much wider radial 
range, the density independent M3Y interactions failed to give a good description 
of the data and the inclusion of an explicit density dependence was found necessary 
\cite{Ko82} to account for the reduction of the attractive strength of the in-medium 
NN interaction that occurs as the density of the nuclear medium increases. Such an 
effect has been shown to be due to the saturation properties of nuclear matter 
and some realistic density dependent versions of the M3Y interaction 
\cite{Kho95,Kho97} have been introduced and used successfully in the folding 
model analysis of the elastic \aN scattering (see the recent review in Ref.~\cite{Kho07}),
and it is natural to expect the same success of this density dependent interaction in the 
study of the elastic \aa scattering. The actual double-folding calculation has shown, 
however, that only the original density independent M3Y interaction can give a reasonable 
description of the elastic \aa scattering at low energies \cite{Avri03}. Such an inconsistency 
of the double-folding model has also been noted earlier in Ref.~\cite{SatLove}, where the
rainbow \aN scattering data implied the inclusion of a realistic density dependence into 
the M3Y interaction, while the elastic \aa scattering data preferred the original density 
independent M3Y interaction. 

In contrast to HI scattering, the elastic \aa scattering data at energies below the reaction 
threshold of 34.7 MeV (in the lab system) can be well described by the real OP only 
\cite{Ali,Buck,SatLove,Avri03}, neglecting the imaginary (absorptive) part of the OP. 
Without the absorption, the elastic \aa data measured accurately over the whole 
observable angular range should be sensitive to the real OP down to small radii where 
the density dependent effects should be substantial due to a high \aa overlap density. 
Thus, the success of the density independent M3Y interaction in the description of the 
considered \aa data indicates likely to a strong depletion of the \aa overlap density that 
suppresses the density dependent effects on the shape and depth of the \aa potential.      

To shed more light on the applicability of the double-folding model in the study of
the \aa scattering at low energies, we have performed in the present work a detailed 
folding model analysis of the available elastic \aa data at energies below the reaction 
threshold. The effects of the density dependence of the NN interaction to the \aa 
potential were studied carefully, based on different assumptions for the \aa overlap density.        

\section{Theoretical formalism}
\label{sec1}

Our microscopic study of the elastic \aa scattering is based on the double-folding 
model (DFM) \cite{Kho00}, which calculates the real OP of the \aa system using 
the ground state density of $^4$He nucleus and an appropriate choice of the 
effective NN interaction. $\alpha$-particle is a unique case when a simple 
Gaussian can reproduce rather well its ground state (g.s) density. Like in numerous  
folding model studies of \aN scattering, we have used in the present DFM calculation 
the Gaussian form for the \al-density suggested by Satchler and Love 
\cite{SatLove}. This \al-density has a RMS radius of 1.461 fm, close to the 
empirical value of $1.47\pm 0.02$ fm that can be deduced from the experimental 
charge density of $^4$He \cite{De74,Dev87}. 

It is straightforward to see that the Gaussian density is readily obtained 
in a simple 4-nucleon model for the \al-particle, where 4 nucleons occupy the 
lowest s${1\over 2}$ harmonic oscillator (h.o.) shell \cite{Kho01}. 
After the spurious center-of-mass (c.m.) component is excluded from 
the 4-nucleon wave function using the prescription of Ref.~\cite{Tas58}, 
the \al-density remains in a Gaussian form but with a modified h.o. range $b$
\begin{equation}
\rho(r)=\frac{4}{\pi^{3/2}b^3}\exp\left(-\frac{r^2}{b^2}\right)\rightarrow
{\rm c.m.\ correction}\rightarrow
\rho(r)=\frac{32}{(3\pi)^{3/2}b^3}\exp\left(-\frac{4r^2}{3b^2}\right). \label{e1a}
\end{equation}         
Therefore, if one assumes $\sqrt{3}~b/2=1.1932$ fm in the \al-density after the
c.m. correction then it turns out to be the same Gaussian as that suggested by
Satchler and Love \cite{SatLove}, which has been used so far in most of the 
folding calculations of the \aa and \aN potentials 
\cite{SatLove,Kho97,Kho07,Ko82,Avri03,Kho01}. In the present work we have 
compared in some cases the results obtained with 
the Gaussian density (\ref{e1a}) with those obtained with the experimental 
\al-density (twice the experimental $^4$He charge density \cite{De74} unfolded 
with the finite size of proton).   

\subsection{Density dependent M3Y interaction}
\label{inter}
\begin{figure}[bht]
 \vspace{4.5cm}\hspace{-1cm}
\includegraphics[angle=0,scale=0.6]{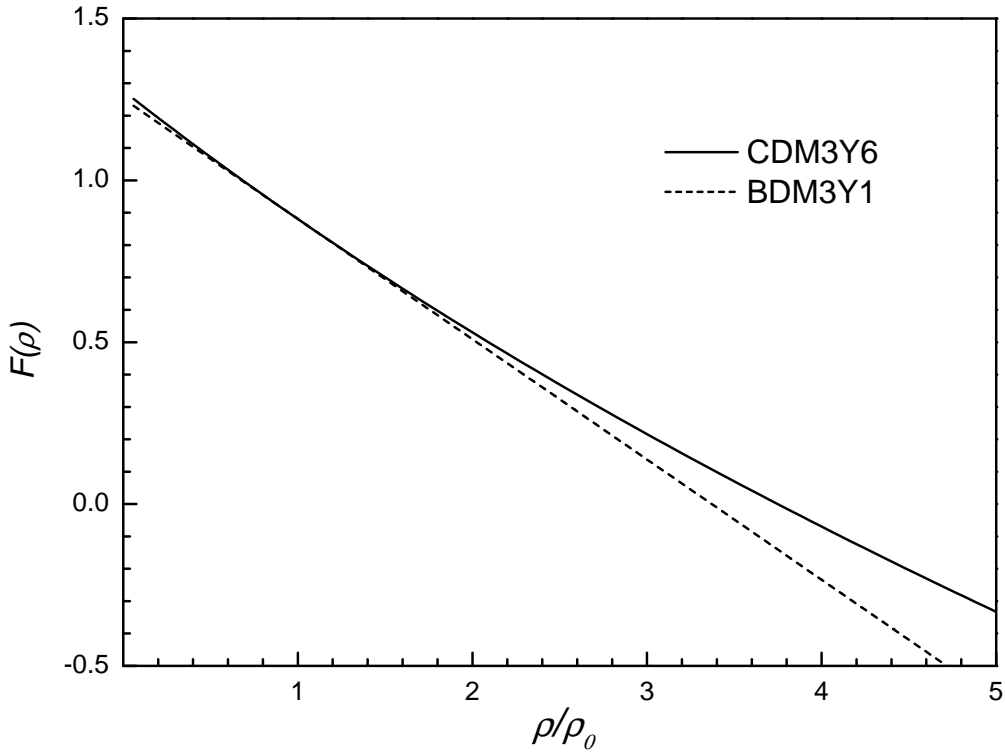}\vspace{-6cm}
\caption{The behavior of the density dependence BDM3Y1 and CDM3Y6 
(see Eq.~(\ref{e3})) of the M3Y-Reid \cite{Be77} and M3Y-Paris \cite{An83} 
interaction, respectively.}
 \label{f0}
\end{figure}

A popular choice of the effective NN interaction for the DFM calculation has
been one of the M3Y interactions that were designed to reproduce the G-matrix
elements of the Reid \cite{Be77} and Paris \cite{An83} NN potentials in an
oscillator basis. Although the original density independent M3Y interaction has 
been used with some success to calculate the real HI optical potential at low energies, 
where the scattering data are sensitive to the real OP only at the surface \cite{SatLove}, 
it failed to account for the nuclear `rainbow' scattering (observed first in the elastic 
\aN scattering and later on in some light heavy-ion systems \cite{Kho07}), when the 
scattering data are sensitive to the real OP over a wider radial range. This has motivated 
the inclusion of an explicit density dependence into the original M3Y interaction 
\cite{Kho95,Kho97,Ko82} to properly account for the reduction of the attractive 
strength of the effective NN interaction occurring at high densities of the nuclear medium 
(see Fig.~\ref{f0}). 

We have chosen for the present study the BDM3Y1 \cite{Kho95} and 
CDM3Y6 \cite{Kho97} density dependent versions of the M3Y interaction that are 
based on the original M3Y-Reid \cite{Be77} and M3Y-Paris \cite{An83} interactions, 
respectively, and parametrized \cite{Kho95,Kho97} as 
\begin{eqnarray}
 v_{\rm D(EX)}(E,\rho,s)=g(E)F(\rho)v_{\rm D(EX)}(s), \label{e2} \\
\mbox{with}\ F(\rho)=C[1+\alpha \exp(-\beta\rho)-\gamma\rho]. \label{e3}
\end{eqnarray}
The radial parts of the direct and exchange parts $v_{\rm D(EX)}(s)$ were
kept unchanged, as derived from the original M3Y interactions, in terms of three
Yukawas \cite{Be77,An83}. The explicit expressions of $v_{\rm D(EX)}(s)$, the 
linear energy dependent factor $g(E)$ and parameters $C,\ \alpha,\ \beta$ and $\gamma$ 
can be found, e.g., in Ref.~\cite{Kho07}. The parameters (\ref{e3}) of the BDM3Y1 
and CDM3Y6 density dependences have been carefully adjusted in the Hartree-Fock 
(HF) calculation to reproduce the saturation of the cold nuclear matter at $\rho=\rho_0$, 
with $\rho_0\approx 0.17$ fm$^{-3}$, and give the nuclear matter incompressibility 
$K\approx 232$ and 252 MeV, respectively \cite{Kho07}. The behavior of the 
density dependent function $F(\rho)$ is shown in Fig.~\ref{f0}, and it has been 
probed quite accurately in the folding model analysis of the refractive  \aN 
scattering, at the densities up to $\rho\approx 2\rho_0$ \cite{Kho97,Kho07}. We note 
that the \al-particle has a very compact density that can be as high as $2\rho_0$ in the 
center \cite{Si76}, and the \emph{static} overlap density of the \aa system may reach 
as much as 4$\rho_0$. Therefore, the use of a \emph{density dependent} NN interaction 
in the folding calculation of the \aa potential should be necessary. 

It should be noted that the BDM3Y1 and CDM3Y6 density dependences have been 
tailored in the HF calculation for a uniform nuclear matter that can be represented 
by a single Fermi sphere in the momentum space. The situation in a \AA collision is
much more complicated, and the momentum distribution of the dinuclear system  is 
a dynamic picture of two Fermi spheres separated by the local relative nucleon 
momentum. From the nuclear matter point of view, a realistic density dependent  
NN interaction for the folding model calculation of the \AA potential (or the dynamic 
simulation of the \AA collision based on a transport model) should be derived 
basically from a Brueckner-Hartree-Fock study of the two slabs of nuclear matter 
separated by different relative nucleon momenta, and at different asymmetries 
of the matter densities of the two slabs. Such an approach has been initiated in the past 
by Tuebingen group \cite{Oht87,Kho90} but remains incomplete. It is, therefore, 
desirable that the issue raised in the present work will give a new motivation for 
such a microscopic study of dinuclear matter.   

\subsection{Double-folding model}
\label{folding}

The generalized DFM of Ref.~\cite{Kho00} was used to evaluate the \aa potential 
from the HF-type matrix elements (\ref{e1}) of the density-dependent interaction 
(\ref{e2})-(\ref{e3}). The (local) direct term is readily evaluated by the standard 
double-folding integration
\begin{equation}
 U_{\rm D}(E,\bm{R})=\int\rho_1(\bm{r}_1)\rho_2(\bm{r}_2)
 v_{\rm D}(E,\rho,s)d^3r_1d^3r_2,\  \bm{s}=\bm{r}_2-\bm{r}_1+\bm{R}.
\label{e4}
\end{equation}
The exchange term in Eq.~(\ref{e1}) is generally nonlocal in the coordinate space, 
but a local form of the exchange potential can be obtained using the local WKB 
approximation \cite{Sat83} for the change in relative motion induced by the 
exchange of spatial coordinates of each interacting nucleon pair \cite{Kho00,Kho07}  
\begin{eqnarray}
U_{\rm EX}(E,\bm{R}) =\int \rho_1(\bm{r}_1,\bm{r}_1+\bm{s})
 \rho_2(\bm{r}_2,\bm{r}_2-\bm{s})v_{\rm EX}(E,\rho,s) \nonumber \\
\times\exp\left({i\bm{K}(\bm{R})\bm{s}}\over{M}\right)d^3r_1d^3r_2.
\label{e5}
\end{eqnarray}
Here $\bm{K}(\bm{R})$ is the local momentum of relative motion determined
from
\begin{equation}
 K^2(\bm{R})={{2\mu}\over{\hbar}^2}[E-U(E,\bm{R})-V_C(\bm{R})],
\label{e6}
\end{equation}
where $\mu$ is the reduced mass, $M=A_1A_2/(A_1+A_2)\equiv 2$ is the recoil
factor or the reduced mass number, $E$ is the scattering energy in the 
center-of-mass frame, $U(E,\bm{R})$ and $V_C(\bm{R})$ are the nuclear 
and Coulomb parts of the \aa potential, respectively. The calculation 
of $U_{\rm EX}$ is done iteratively using the explicit expression of the 
nonlocal density matrix given by the h.o. wave functions of nucleons bound in 
the two \al-particle \cite{Kho01}. To validate the folding model prediction for the 
\AA potential, it is important to discuss the treatment of the dinuclear (overlap) 
density embedded in the density dependence (\ref{e3}) of the M3Y interaction. 
In the present paper we consider three approximations for the \aa overlap density  

\subsection*{Frozen density approximation}
We recall that the DFM generates the first-order term of the microscopic OP in 
the Feshbach's scheme \cite{Fe92}, used in the OM equation to obtain the 
relative-motion wave function of the two colliding nuclei being in their ground states. 
Given the antisymmetrization effects accurately taken into account via the exchange 
term (\ref{e5}), a reasonable approximation for the total density $\rho$ of the two 
overlapping nuclei is the sum of the two g.s. densities. In the calculation of the direct 
potential (\ref{e4}) the overlap density $\rho$ in $F(E,\rho)$ is taken as the sum 
of the two $\alpha$ densities at the position of each nucleon
\begin{equation}
  \rho=\rho_1 (\bm{r}_1) + \rho_2(\bm{r}_2).
 \label{e7}
\end{equation}
The assumption (\ref{e7}) was widely adopted in the DFM calculations with 
the density dependent NN interaction
\cite{SatLove,Kho95,Kho97,Kho07,Ko82,Avri03,Kho01} because it
allows an explicit separation of variables in the three-dimensional
integral (\ref{e4}). In evaluating the exchange potential (\ref{e5}), the
overlap density in $F(\rho)$ is taken as the sum of the two $\alpha$ densities
at the midpoint between the two nucleons being exchanged \cite{Kho00}
\begin{equation}
 \rho=\rho_{1}\left(\bm{r}_1+\frac{\bm{s}}{2}\right)+
 \rho_{2}\left(\bm{r}_2-\frac{\bm{s}}{2}\right). \label{e7x}
\end{equation}
The approximation (\ref{e7})-(\ref{e7x}), dubbed as \emph{frozen density approximation} 
(FDA), has been used in most of the DFM calculations of the \AA potential using a density 
dependent NN interaction when the energy is not too low. Any density rearrangement 
that might happen during the collision would lead to the nuclear states different from the 
ground states, and thus contribute to higher-order dynamic polarization potential in the
Feshbach's scheme \cite{Fe92}. The FDA reproduces very well the observed
reduction of the attractive strength of the real \AA OP at small distances implied,
in particular, by the refractive \aN scattering data \cite{Kho01,Kho07}. The use of FDA 
has also been shown in the recent DFM calculations of the \AA OP at medium energies by 
RIKEN-Osaka group \cite{Fur09,Fur10,Fur12}, using a realistic G-matrix interaction, as 
the most suitable for the \AA overlap density. 

In the \aa case, the static FDA gives the overlap density reaching $3\sim 4\rho_0$ 
at the smallest distances, and the considered density dependence of the M3Y 
interaction has not been tested at such a high density. Moreover, it is also 
questionable if a simple geometrical overlap of the two g.s. densities implied by 
the FDA is still a reasonable approximation for the density dependence (\ref{e3}) 
at very low energies. As noted above, a realistic density dependence of the 
effective NN interaction for the DFM calculation of the \AA potential should be, 
in general, constructed from a microscopic study of the dinuclear matter at different 
momentum separations and density asymmetries. In the momentum space, the distance 
separating the two colliding slabs of nuclear matter becomes small at low energies 
and the Pauli blocking should play a very crucial role in shaping the density 
dependence of the NN interaction in the dinuclear medium.  

In addition to the FDA, other approximations have also been used for the overlap
density in the DFM calculation like, e.g., the geometric or arithmetic averages of the 
two g.s. densities adopted \cite{Tr00} in the folding calculations using the JLM 
density dependent interaction \cite{Je77}. 

\subsection*{Fully antisymmetrized density of the \aa system}
\aa system is a very special case where one could check the validity of FDA for the 
overlap density by estimating explicitly the \aa density from the fully antisymmetrized
total wave function of 8 nucleons bound in the two \al\ clusters. Using the microscopic 
cluster model suggested by Brink for $^8$Be resonance \cite{Hor10}, the total density 
of the \aa system at a given distance $R$ between the centres of mass of the two 
\al-particles can be determined as
\begin{equation}
\rho(\bm{r},R)=\langle \Psi(R)| \sum_{i=1}^{8} 
 \delta(\bm{r}-\bm{r}_i) | \Psi(R)\rangle, \label{e8}    
\end{equation}
where the total wave function $\Psi(R)$ of the system is determined as a Slater
determinant of the single-particle wave functions of 8 nucleons bound
in the two \al\ clusters    
\begin{equation}
\Psi(R)={\mathcal A} \{\psi_{\alpha_1}\times \psi_{\alpha_2}\}. \label{e9}    
\end{equation} 
Here ${\mathcal A}$ is the antisymmetrizer and $\psi_{\alpha_{1(2)}}$  
is the antisymmetrized wave function built upon the single-particle wave 
functions of 4 nucleons bound in each \al\ cluster. The single-particle wave functions
are the same h.o. wave functions as those used to calculate the \al-density used in 
the present folding calculation. The details of the \aa density calculation 
(\ref{e8})-(\ref{e9}) are given in the Appendix. We denote hereafter the use of the 
\emph{antisymmetrized overlap density} in the density dependence $F(\rho)$
of the NN interaction as the AOD approximation.    

The AOD procedure (see the next session and appendix) substantially changes 
the radial shape of the \aa overlap density at small distances, and it is no more 
a direct sum of the two \al-densities. As can be seen from Eq.~(\ref{eq8}), the fully 
antisymmetrized density contains two remnants of the original \al-densities and 
an "interference" term that arises from the antisymmetrization of the 8-nucleon 
wave function. As a result, the simple geometrical overlap of the two \al-densities 
implied by the FDA does not hold any more if the full antisymmetrization is taken 
into account. We will consider in the next session how the AOD procedure affects 
the density dependence (\ref{e3}) of the M3Y interaction, and whether this 
helps to clarify the inconsistency of the DFM discussed in the introduction.  
 
\subsection*{Dynamic distortion of the FDA by Pauli blocking}
\begin{figure}[bht]
 \vspace{4.5cm}\hspace{-3.5cm}
\includegraphics[angle=0,scale=0.7]{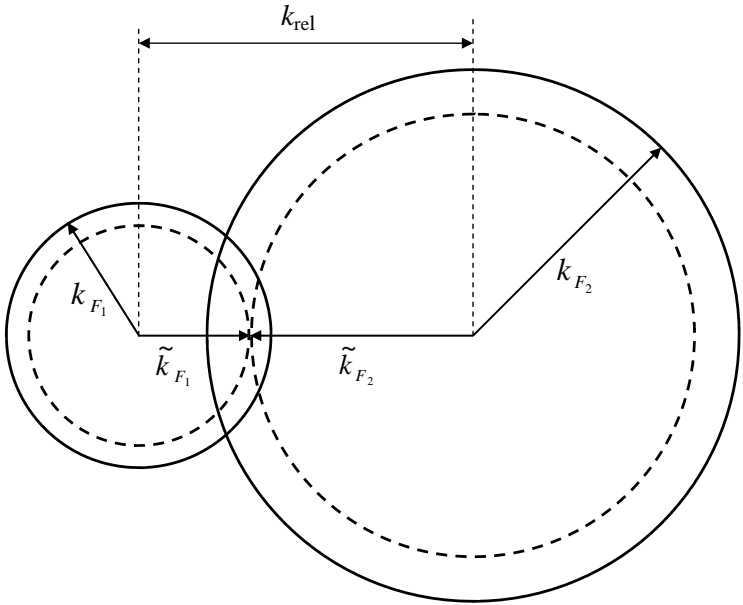}\vspace{-9cm}
\caption{The dynamic Pauli distortion of the two Fermi spheres representing the local 
densities of the two colliding nuclei in the momentum space.}
 \label{f1}
\end{figure}

The AOD treatment is still a \emph{static} approximation for the density 
dependence (\ref{e3}) of the M3Y interaction. To estimate the \emph{dynamic} 
distortion of the density dependence by the Pauli blocking, we refer to the nuclear 
matter approach to study \AA OP by Tuebingen group \cite{Oht87,Kho90}, 
which has been improved and further developed by Soubbotin {\it et al.} \cite{Sou01}. 
Namely, a \AA collision can be locally considered as a collision of two slabs 
of nuclear matter whose densities are the local densities of the target and projectile. 
The momentum distribution of the dinuclear system with the given local densities 
$\rho_1$ and $\rho_2$ is represented by the two Fermi spheres with radii 
$k_{F_1}=(1.5\pi^2\rho_1)^{1/3}$ and $k_{F_2}=(1.5\pi^2\rho_2)^{1/3}$, and
their centres separated by the average momentum of the nucleon relative motion 
$k_{\rm rel}$ (see Fig.~\ref{f1}). 
The original approach by Tuebingen group \cite{Oht87,Kho90} has used 
$k_{\rm rel}=K_\infty/M$, where $K_\infty$ is derived from Eq.~(\ref{e6}) at 
$R\to\infty$ and $M$ is the recoil factor in Eq.~(\ref{e5}). However, the OP obtained 
in this nuclear matter approach has been found later to be out of the global systematics 
established for the \AA OP \cite{Kho07,Bra97}. In the present study we have used 
the local relative-motion momentum of nucleon $k_{\rm rel}(R)=K(R)/M$, with $K(R)$ 
determined self-consistently from the double-folded potential by Eq.~(\ref{e6}). 
Such a treatment directly links the momentum distribution of the dinuclear density 
to the potential strength at each internuclear separation $R$, and it allowed to explain 
\cite{Sou01} the deep mean-field-type potential established by the global systematics 
\cite{Bra97} or predicted by the DFM \cite{Kho07}. The Pauli blocking forbids the 
overlap of the two Fermi spheres in the momentum space, and the shapes of the two 
Fermi spheres must be modified wherever $k_{F_1}+k_{F_2}>k_{\rm rel}(R)$. 
In the nuclear matter approaches \cite{Oht87,Kho90,Sou01} the OP between  
two nuclei separated by a distance $R$ is determined as the difference of the total 
energy of the dinuclear system at $R$ from that at infinite distance, based on the 
energy density formalism. Within this formalism, the total density ($\rho_1+\rho_2$) 
must be unchanged and the Pauli distortion results, therefore, on a \emph{non-spherical} 
shape of each distorted Fermi sphere (see, e.g., Fig.~1 in Ref.~\cite{Sou01}). 
The non-spherical shapes of the two density distributions in the momentum space imply, 
however, that the two nuclei are no more in their ground states but in some (Pauli) excited 
states. As a result, the \emph{Pauli distorted} total energy does not determine just the 
relative motion of the two nuclei being in their ground states, but the motion 
of a wave packet that includes also excited states. Such a wave packet is, in general, 
not appropriate for the description of elastic scattering \cite{Bra97}. 

To remain in the framework of the standard DFM \cite{Kho00,Kho97} without any 
hybrid coupling to the energy density formalism, we have suggested in the present 
work a procedure to improve the DFM by taking into account only the Pauli distortion 
of the density dependence (\ref{e3}) of the M3Y interaction. The densities entering 
the double-folding integration (\ref{e4})-(\ref{e5}) remain unchanged, so that all the 
pair-wise interactions (\ref{e1}) between the projectile- and target nucleons are fully 
taken into account. Thus, the density dependent strength of the M3Y interaction is 
not static as adopted in the FDA, but is dynamically modified by the Pauli blocking 
at the small internuclear distances $R$. Whenever $k_{F_1}+k_{F_2}>k_{\rm rel}(R)$, 
the corresponding local densities are reduced ($\rho_{1,2}\to\tilde \rho_{1,2}$) so 
that the radii of the two reduced Fermi spheres satisfy relation   
$\tilde k_{F_1}+\tilde k_{F_2}=k_{\rm rel}(R)$ that is allowed by the 
Pauli principle (see Fig.~\ref{f1}). The total density in (\ref{e3}) is reduced 
by such a ``shrinkage" of the two local densities, but the projectile-target asymmetry 
$\chi$ is kept unchanged 
\begin{equation}
 \chi=\frac{\rho_1}{\rho_1+\rho_2}=\frac{k^3_{F_1}}{k^3_{F_1}+k^3_{F_2}}
 \equiv\frac{\tilde\rho_1}{\tilde\rho_1+\tilde\rho_2}=\frac{\tilde k^3_{F_1}}
 {\tilde k^3_{F_1}+\tilde k^3_{F_2}}. \label{e10}    
\end{equation} 
Using condition (\ref{e10}) it is straightforward to obtain the reduced radii of the 
two distorted Fermi spheres as 
\begin{equation}
 \tilde k_{F_1}=\frac{k_{\rm rel}}{1+X_\chi},\ 
 \tilde k_{F_2}=\frac{k_{\rm rel} X_\chi}{1+X_\chi},\ {\rm with}\ 
 X_\chi=\left(\frac{1-\chi}{\chi}\right)^{\frac{1}{3}}. \label{e11}
\end{equation} 
In difference from the nuclear matter approaches \cite{Oht87,Kho90,Sou01}, the two 
Fermi spheres distorted by the Pauli blocking remain spherical in this case (shown by 
dashed lines in Fig.~\ref{f1}). We denote hereafter the use of the overlap 
density modified by the \emph{dynamic Pauli distortion} 
($\rho=\tilde \rho_1+\tilde \rho_2$)  in Eq.~(\ref{e3}) as the DPD approximation. 

We stress again that, similar to the AOD procedure, the Pauli distorted local 
densities $\tilde \rho_{1,2}$ are used only to determine the overlap density used in 
the density dependence (\ref{e3}) of the M3Y interaction. The local and nonlocal 
nuclear densities entering the direct and exchange folding integrals (\ref{e4})-(\ref{e5}) 
remain the original g.s. densities of the projectile and target, so that the DFM still 
determines the \aa  potential in the first order of the Feshbach's theory \cite{Fe92}. 
As can be seen from the discussion in the next section, the DPD procedure strongly 
reduces the dinuclear matter density ($\rho=\tilde\rho_1+\tilde\rho_2$) entering the 
density dependence (\ref{e3}) of the M3Y interaction and helps to explain the 
observed depletion of the overlap \aa density. On the other hand, the distorted total 
density is no more a direct sum of the two g.s. densities used in the double-folding 
integration (\ref{e4})-(\ref{e5}), and this actually shows the breakdown of the FDA 
in the DFM description of the \aa scattering at low energies.    

\section{Treatment of the densities and \aa potential}
\label{sec2}
The elastic \aa scattering data are widely available at energies ranging from as low 
as 0.6 MeV up to GeV region. In the present study we have considered only the  
data measured at the laboratory energies below the reaction threshold of 34.7 MeV, 
where the imaginary (absorptive) part of the OP is negligible and the elastic \aa data
can be well described by the real OP only. Different versions of the folded 
\aa potential (\ref{e1}) were used in the present OM analysis of elastic \aa data at 
energies of $E_\alpha=3$  to 29.5 MeV \cite{Hey56,Tom63,Nil56,Ste53,Chi74}. 
To obtain the total OP, a Coulomb potential $V_C(R)$ needs to be added to the 
folded potential (\ref{e1}). In the OM studies of elastic \AA scattering, $V_C(R)$ is 
often chosen as the Coulomb potential between a point charge and a uniform charge 
distribution of the radius $R_{C}$. A more realistic choice of $V_C(R)$ for the \AA 
scattering is the Coulomb potential generated by double-folding two uniform charge 
distributions with radii $R_{C_1}=r_{C_1}A_1^{1/3}$ and 
$R_{C_2}=r_{C_2}A_2^{1/3}$ for which an analytic expression is available \cite{De75}. 
Both prescriptions give the correct asymptotic expression for the Coulomb potential,
\begin{equation}
 V_C(R)\to Z_1Z_2e^2/R,
\label{e12}
\end{equation}
but differ at small radii where the two nuclei overlap \cite{Bra97}. We have used in the
present work the folded Coulomb potential \cite{De75} for the \aa system, with the radii 
$r_{C_1}=r_{C_2}=1.36$ fm. All OM calculations were done
using the direct reaction code PTOLEMY \cite{Rh80}.

\begin{figure}[bht]
\vspace{3cm}\hspace{0cm}
\includegraphics[angle=0,scale=0.65]{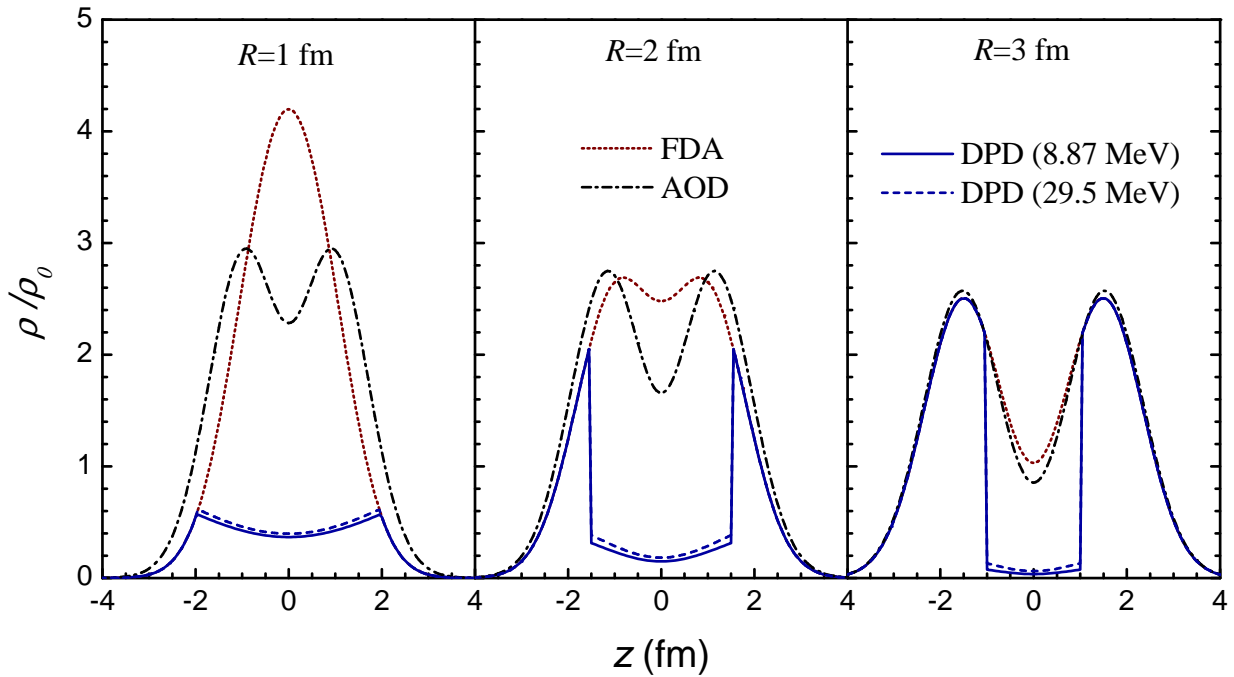}\vspace{-7cm}
\caption{The overlap \aa density at different distances $R$ between 
the centres of the two \al-particles, given by different approximations for the density 
dependence of the CDM3Y6 interaction. The $z$-axis is aligned along the beam 
direction, with the origin at the center of mass of the \aa system.} \label{f2}
\end{figure}

\begin{figure}[bht]
 \vspace{3cm}\hspace{-1cm}
\includegraphics[angle=0,scale=0.6]{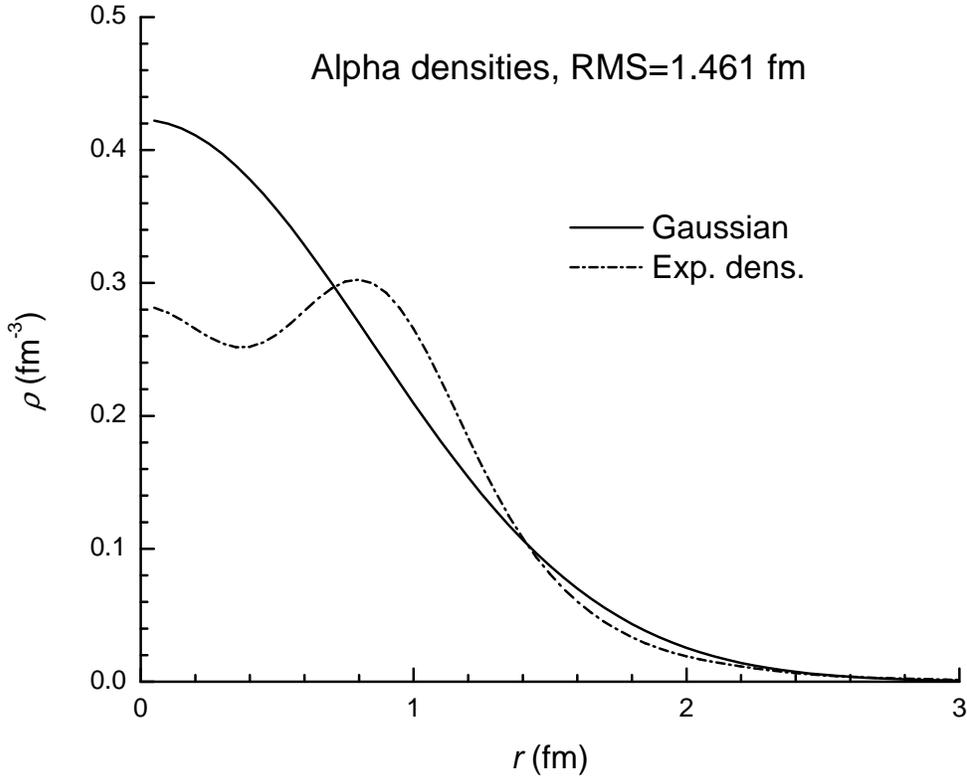}\vspace{-4cm}
\caption{The Gaussian \al-density (\ref{e1a}) and the experimental \al-density 
(twice the experimental $^4$He charge density \cite{De74} unfolded with the finite 
size of proton using the prescription of Ref.~\cite{SatLove}). Both densities
have the same RMS radius of 1.461 fm}.   
 \label{f2k}
\end{figure}
\begin{figure}[bht]
\vspace{3.5cm}\hspace{0cm}
\includegraphics[angle=0,scale=0.6]{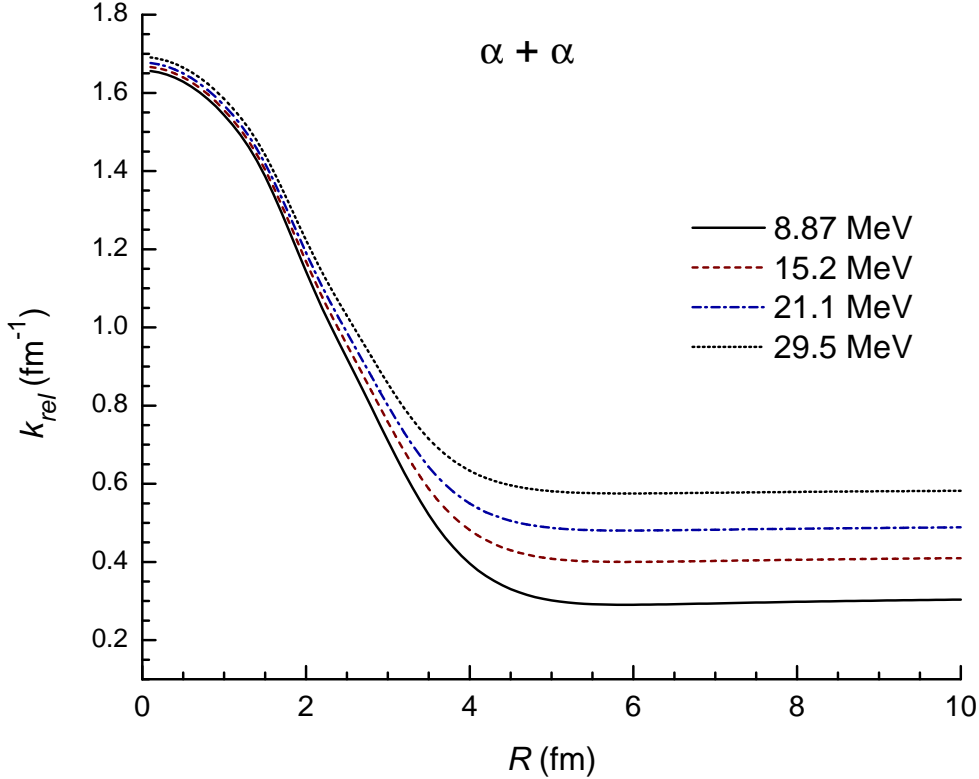}\vspace{-4.5cm}
\caption{The relative nucleon momentum $k_{\rm rel}=K(R)/M$ 
determined at different energies from the CDM3Y6 folded potential given by 
the DPD approximation for the density dependence.} \label{f3}
\end{figure}

As noted above, the use of the density dependent NN interaction in the DFM analysis 
of the elastic \aa scattering at low energies was not successful \cite{Avri03}
when the FDA was used for the overlap density in the density dependence (\ref{e3}). 
The folded potential is usually too shallow in this case and fails to account 
for the elastic \aa scattering data. The DFM can give a reasonable prediction 
of the \aa potential only when the density dependence is neglected, i.e., 
to put $F(E,\rho)=1$ in Eq.~(\ref{e3}) and use the original density 
independent M3Y interactions \cite{Be77,An83}. It is trivial to find out that this 
effect is due to a very high density profile of the $^4$He nucleus. It is well 
established from the electron scattering data \cite{Si76} that the matter density 
of the $\alpha$-particle is $\rho\simeq 2\rho_0$ around its center. As a result, 
the total density of the two $\alpha$-particles overlapping each other is reaching 
as much as 4$\rho_0$ in the FDA, and the folded \aa potential becomes too shallow 
at small radii due to a strong repulsion caused by the steady decrease of $F(\rho)$ 
at high densities (see Fig.~\ref{f0}). The use of a more sophisticated, fully 
antisymmetrized overlap density (see Appendix) does not solve the problem. 
Although the overlap density is reduced in the center of the \aa system by the full 
antisymmetrization (see Fig.~\ref{f2}), the folded potentials obtained with the AOD 
approximation are even slightly shallower than the folded potentials given by the 
FDA (see Fig.~\ref{f3}) due to a higher density given by AOD at the sub-surface 
region. It becomes obvious that at the low energies, the \emph{static} overlap 
density of the \aa system cannot be used in the density dependence (\ref{e3}) 
of the M3Y interaction, even when the antisymmetrization of the overlap density 
is taken into account exactly.

\begin{figure}[bht]
 \vspace{3.5cm}\hspace{0cm}
\includegraphics[angle=0,scale=0.6]{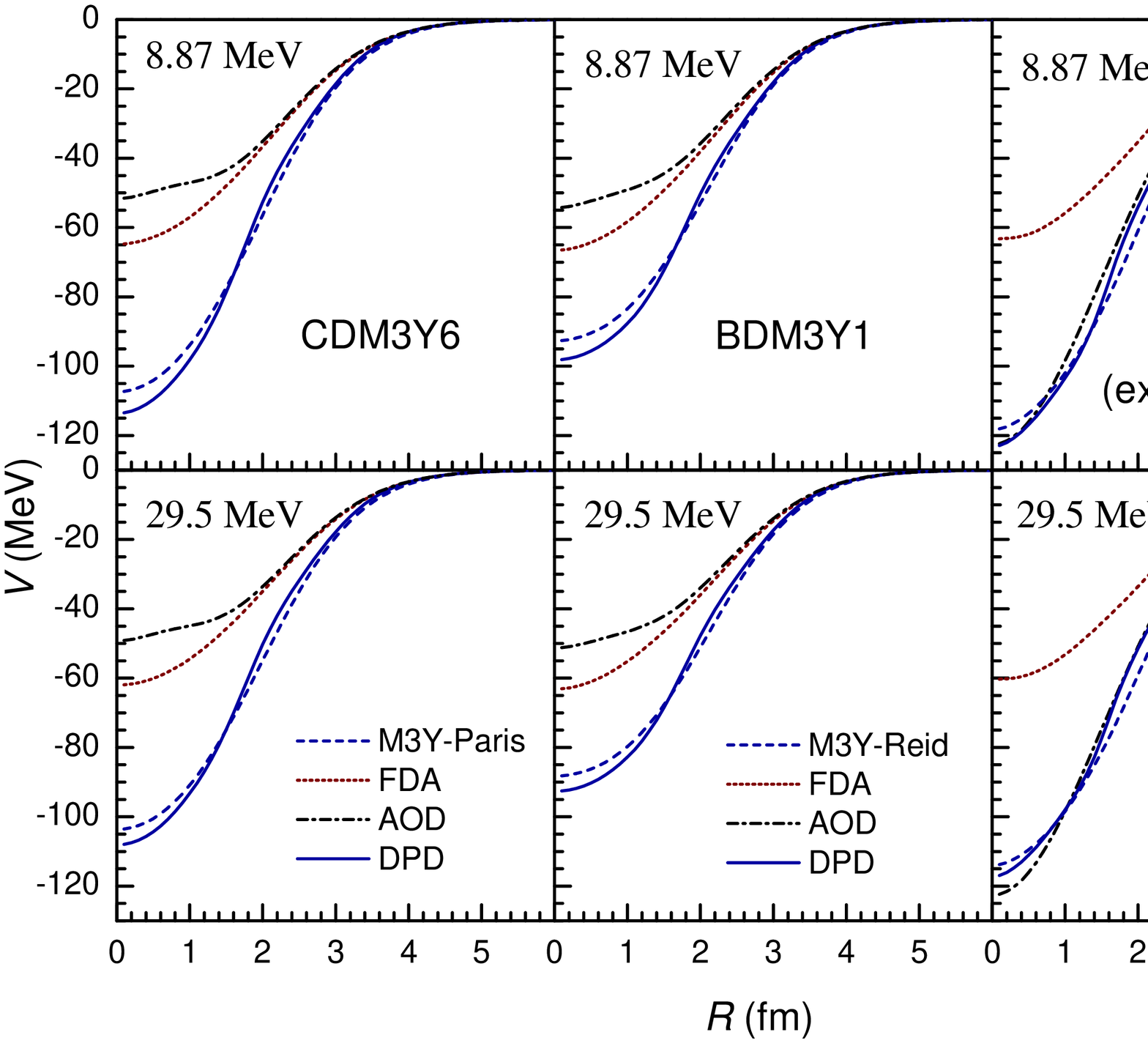}\vspace{-4cm}
\caption{The (unrenormalized) folded \aa potentials at the incident energies 
$E_\alpha=8.87$ and 29.5 MeV given by different approximations for the density 
dependence of the CDM3Y6 and BDM3Y1 interactions, in comparison with the 
(energy independent) Gaussian potentials (\ref{e13}) proposed by Buck {\em et al.} 
\cite{Buck}.} \label{f4}
\end{figure}
In a \emph{dynamic} picture discussed in Sec.~\ref{sec1}, the Pauli blocking does 
not allow the overlap of the two Fermi spheres representing the local densities of the 
two \al-particles. At the considered energies, the distance between the two Fermi 
spheres, i.e., the nucleon relative momentum $k_{\rm rel}(R)$ is quite small and 
reaches its maximum of $1.65\sim 1.7$ fm$^{-1}$ at the smallest radii $R$  
(see Fig.~\ref{f3}). With the maximal radius $k_F$ of each Fermi sphere lying 
around $1.70\sim 1.75$ fm$^{-1}$ (as given by the empirical \al\ density \cite{Si76}), 
the maximal $k_{\rm rel}(R)$ value required by the Pauli blocking 
must be around $3.4\sim 3.5$ fm$^{-1}$ that is much larger than its maximum at the
small distances. As a result, the two local densities entering the density dependence 
$F(\rho)$ should be strongly reduced by the dynamic Pauli distortion, 
$\tilde\rho_{1,2}<\rho_{1,2}$, as illustrated in Fig.~\ref{f1}. To implement the 
DPD procedure we check, at each distance $R$ between the centres of the two 
\al-particles, all possible overlaps of the two \al-densities in the coordinate space 
$\rho (\bm{r},R)=\rho_1(\bm{r},R) + \rho_2(\bm{r},R)$, where $\bm{r}$ is the radius
vector in the coordinate system with the origin at the center of mass and the $z$-axis
lying along the beam direction as shown in Fig.~\ref{f2}. To prevent the 
self-contraction of a Fermi sphere having $k_F>k_{\rm rel}(R)$ in the region where 
only one \al-density is dominated and the other \al-density is negligible, we have 
applied the DPD treatment only at the locations where both \al-particles have density 
$\rho\geqslant 0.005$ fm$^{-3}$. At such a location the relation 
\begin{equation}
 k_{F_1}+k_{F_2}\leqslant k_{\rm rel}, \label{e12a}
\end{equation}
is being checked and $\rho_{1,2}(\bm{r},R)$ are replaced  
by $\tilde\rho_{1,2}(\bm{r},R)$ using the prescription (\ref{e11}) wherever the 
relation (\ref{e12a}) is not fulfilled. Such a DPD procedure is done iteratively until 
the radii of the two Fermi spheres always satisfy the relation (\ref{e12a}). A lower 
limit of the $\alpha$-density automatically stops the DPD treatment at large distances 
$R$ where the two \al-particles are well separated and do not overlap in the 
coordinate space. 
This iterative DPD procedure consumes most of the CPU time in our dynamical 
double-folding calculation, which is about 3 orders of magnitude longer than that 
needed for the standard DFM calculation using the static overlap density given 
by the FDA. From the results shown in Fig.~\ref{f2} one can see a substantial 
depletion of the central density at small distances resulted from the DPD treatment 
of the overlap density. At large distances ($R\geqslant 3$ fm), the overlap of the 
two \al-particles in the coordinate space becomes less significant and the depletion 
of the overlap density occurs in a much smaller central spot. However, the 
density inside this central spot decreases very quickly, due to a fast drop of the 
$k_{\rm rel}(R)$ values in the radial region 2 fm $\lesssim R\lesssim$ 4 fm  
(see Fig.~\ref{f3}). Outside the central spot of the depleted density, the three 
approaches give about the same overlap density. 

The folded \aa potentials at $E_\alpha=8.87$ and 29.5 MeV given by the three 
approximations for the density dependence (\ref{e2}-\ref{e3}) of the CDM3Y6 and BDM3Y1 
interactions are shown in Fig.~\ref{f4}. A strong depletion of the central density 
by the DPD procedure leads readily to a much deeper (density dependent) potential
that is comparable in strength and shape with the (density independent) 
M3Y-Paris and M3Y-Reid folded potentials. The folded potentials given by 
the M3Y-Reid interaction and its density dependent versions turned out to be 
slightly shallower than those given by the M3Y-Paris interaction. To be sure that the
effects discussed here are not associated with a particular choice of the \al-density, 
we have used in our DFM calculation also the experimental \al-density, taken as twice 
the experimental $^4$He charge density \cite{De74} unfolded with the finite 
size of proton using the prescription of Ref.~\cite{SatLove}. The radial shapes of the
two \al-densities are shown in Fig.~\ref{f2k} and some difference between them 
can be seen at small radii, although the two densities give the same RMS radius of 1.46 fm. 
The M3Y-Paris and CDM3Y6 folded potentials given by the experimental \al-density 
are plotted in the right panel of Fig.~\ref{f4}, and the same drastic difference has been
found in the results given by the FDA and DPD treatments of the density dependence
of the M3Y-Paris interaction. In general, the folded potentials given by the experimental 
\al-density are somewhat deeper than those given by the Gaussian density (\ref{e1a}). 

\section{Description of the elastic \aa scattering}
As mentioned above, the resonating group method \cite{Buck, Ho91} is a rigorous 
approach that takes into account the full antisymmetrization of the total wave function 
of the \aa system. It has been shown by Buck {\em et al.} \cite{Buck} that about 
the same elastic \aa phase shifts as well as other results of the full RGM 
calculation could be reproduced with an energy independent Gaussian potential 
\begin{equation}
V(R)=-V_0\exp(-\beta R^2), \label{e13}
\end{equation}   
where $V_0=122.6225$ MeV and $\beta=0.22$ fm$^{-2}$.  
The local potential (\ref{e13}) has been used quite successfully as the \al-\al\ 
interaction in some cluster-folding calculations of the real OP between two 
\al-cluster nuclei \cite{Ela01,Kar06}. The folded potentials at $E_\alpha=8.87$ and 
29.5 MeV given by the experimental \al-density are compared with the Gaussian 
potential (\ref{e13}) in the right panel of Fig.~\ref{f4}, and one can see that the 
density independent M3Y-Paris and CDM3Y6 (DPD) potentials are quite close 
to the Gaussian potential at both energies. The folded CDM3Y6 (FDA) potential 
is too shallow and strongly differs from the Gaussian and CDM3Y6 (DPD) potentials 
at small radii. 

\begin{figure}[bht]
\vspace{3.5cm}\hspace{0cm}
\includegraphics[angle=0,scale=0.6]{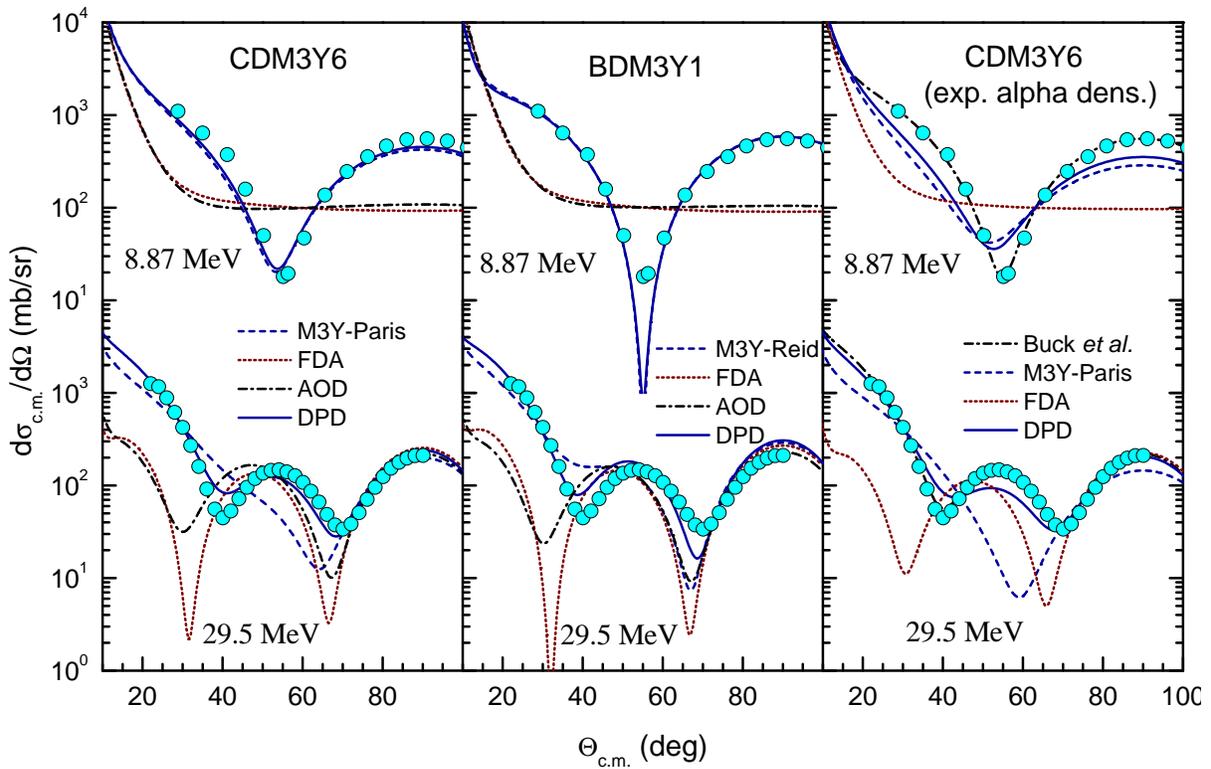}\vspace{-5cm}
\caption{The OM description of the elastic \aa data measured at 
$E_\alpha=8.87$ and 29.5 MeV given by the \emph{unrenormalized} folded potentials 
obtained with different approximations for the density dependence of the CDM3Y6 
(left panel) and BDM3Y1 (middle panel) interactions. The results given by the CDM3Y6 
folded potentials obtained with the experimental \al-density are compared (right panel) 
with those given by the Gaussian potential of Buck {\em et al.} \cite{Buck}.} \label{f5}
\end{figure}
The OM description of the elastic \aa scattering data at $E_\alpha=8.87$ 
and 29.5 MeV given by the folded potentials discussed in Fig.~\ref{f4} are
shown in Fig.~\ref{f5}. From the left and middle panels of Fig.~\ref{f5} one 
can see that the folded potentials obtained with the FDA and AOD approximations
for the overlap density fail badly to account for the measured scattering cross sections. 
A reasonable description of the data is given by the folded potential obtained with the 
density independent M3Y interaction as found earlier in Ref.~\cite{Avri03}, but a 
much better description is given by the folded potential obtained with the density 
dependent CDM3Y6 or BDM3Y1 interaction and the DPD treatment of the density 
dependence. The OM results given by the folded potentials obtained with experimental 
\al-density are compared with those given by the Gaussian potential (\ref{e13}) 
of Buck {\em et al.} \cite{Buck} in the right panel of Fig.~\ref{f5}. For the consistency, 
the Gaussian potential (\ref{e13}) has been used with the Coulomb potential 
$V_C(R)=4e^2{\rm erf}(aR)/R$, with the $a$ value taken from Ref.~\cite{Buck}. 
From two versions of the folded potentials obtained with the DPD approximation, 
the CDM3Y6 potential (based on the M3Y-Paris interaction) accounts for the data 
better than the BDM3Y1 potential (based on the M3Y-Reid interaction). The 
Gaussian potential of Buck {\em et al.} gives about the same good description 
of the data as that given by the CDM3Y6 potential obtained with the DPD 
approximation using the Gaussian density (\ref{e1a}) for the \al-particle.  

\begin{figure}[bht]
 \vspace{3.5cm}\hspace{0cm}
\includegraphics[angle=0,scale=0.6]{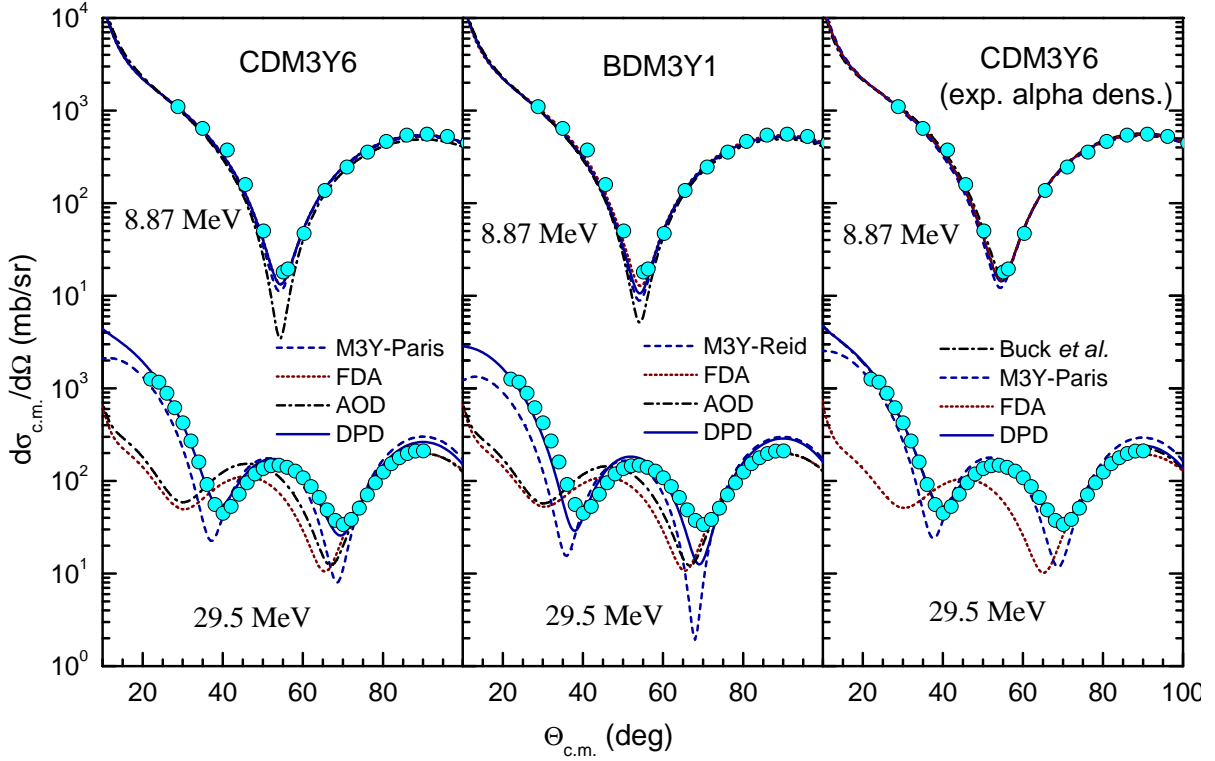}\vspace{-5cm}
\caption{The same as Fig.~\ref{f5} but given by the 
\emph{renormalized} folded and Gaussian potentials. See the renormalization 
coefficient $N$ and other properties of the \aa optical potential in Table~\ref{t1}.} 
 \label{f6}
\end{figure}
\begin{figure}[bht]
 \vspace{0cm}\hspace{0cm}
\includegraphics[angle=0,scale=0.6]{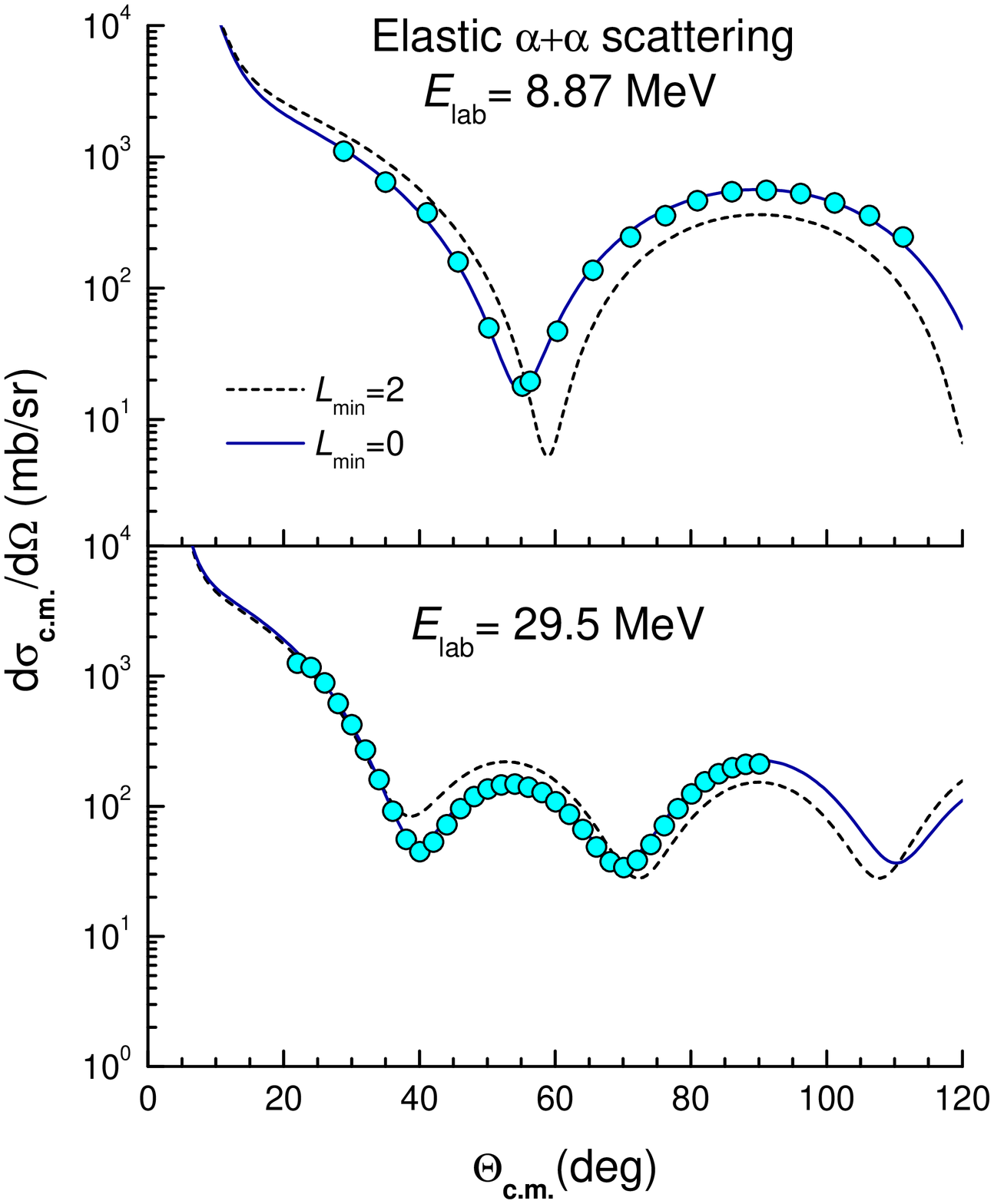}\vspace{-1cm}
\caption{The OM description of the elastic \aa data measured 
at $E_\alpha=8.87$ and 29.5 MeV given by the Gaussian potential of Buck 
{\em et al.} \cite{Buck}, with the lowest partial wave $L_{\rm min}=0$ and 2.} 
 \label{f6k}
\end{figure}
At the considered energies, when only the elastic channel is open and         
there is no coupling to other nonelastic channels, all the \aa potentials under 
study should be used as given by the model, without any further renormalization 
of the potential strength. However, the \aa potentials used in the present 
study are based on certain approximations and an adjustment (or renormalization) 
of the potential strength to the best OM fit of the data should be helpful in 
testing the potential model. We have made, therefore, also the OM calculation 
with different \aa potentials, renormalizing the potential strength to obtain 
the best OM fit of the elastic data in each case. It is natural to expect 
that a realistic model for the \aa potential should give its best OM description 
of the data with a renormalization coefficient $N$ close to unity. The model 
becomes less meaningful if $N$ strongly deviates from unity 
\cite{SatLove,Kho00,Kho97}. The OM results given by the renormalized \aa 
potentials are compared with the data in Fig.~\ref{f6}, and the main 
properties of the potential are given in Table~\ref{t1}. 
\begin{table}\centering
\caption{The properties of the \aa potentials at $E_\alpha=8.87$ and 29.5 MeV
that give the OM results shown in Fig.~\ref{f6}. $N$ is the renormalization 
coefficient of the potential found from the $\chi^2$-fit of the OM results to the 
elastic data. The folded potentials were obtained with the two choices of the 
\al-density, the Gaussian (\ref{e1a}) and experimental density (see Fig.~\ref{f2k}).
$\chi^2$ value is per data point and obtained with uniform 10\% errors.} 
\vspace*{1cm}
\begin{tabular}{|c|c|c|c|c|c|c|c|} \hline
$E_\alpha $& \al-density & Potential & $N$ & $-J_V$ & 
  $\langle r^2 \rangle ^{1/2}_V$ & $\chi^2$ \\ 
(MeV) & &  &  & (MeV~fm$^3$) & (fm) &   \\ \hline
8.87  & - & Buck {\em et al.} & 0.996 & 412.0 & 2.611 & 0.6 \\
  & (\ref{e1a}) & M3Y-Paris& 0.966 & 438.3 & 2.652 & 2.8  \\
  & exp & M3Y-Paris& 0.902 & 435.2 & 2.645 & 2.2  \\
  & (\ref{e1a}) & CDM3Y6 (DPD)  & 0.971 & 419.7 & 2.618 & 1.7  \\
  & exp & CDM3Y6 (DPD)  & 0.935 & 415.1 & 2.615 & 1.2  \\
  & (\ref{e1a}) & CDM3Y6 (FDA)  & 0.587 & 187.5 & 2.777 & 1.5  \\
  & exp & CDM3Y6 (FDA)  & 0.611 & 187.8 & 2.796 & 1.1  \\
  & (\ref{e1a}) & CDM3Y6 (AOD)  & 0.628 & 190.1 & 2.811 & 8.8  \\ \hline
29.5  & - & Buck {\em et al.} & 0.990 & 409.6 & 2.611 & 0.6  \\
  & (\ref{e1a}) & M3Y-Paris  & 0.815 & 361.4 & 2.661 & 17.4  \\
  & exp & M3Y-Paris  & 0.784 & 368.2 & 2.652 & 13.2  \\
  & (\ref{e1a}) & CDM3Y6 (DPD)  & 0.958 & 396.5 & 2.627 & 3.3  \\
  & exp & CDM3Y6 (DPD)  & 0.949 & 401.8 & 2.621 & 1.2  \\
  & (\ref{e1a}) & CDM3Y6 (FDA)  & 0.889 & 272.7 & 2.785 & 34.8  \\
  & exp & CDM3Y6 (FDA)  & 0.922 & 271.0 & 2.805 & 34.5  \\
  & (\ref{e1a}) & CDM3Y6 (AOD)  & 0.959 & 279.5 & 2.820 & 46.0  \\ \hline
\end{tabular} \label{t1} 
\end{table}
\begin{figure}[bht]
 \vspace{3.5cm}\hspace{0cm}
\includegraphics[angle=0,scale=0.65]{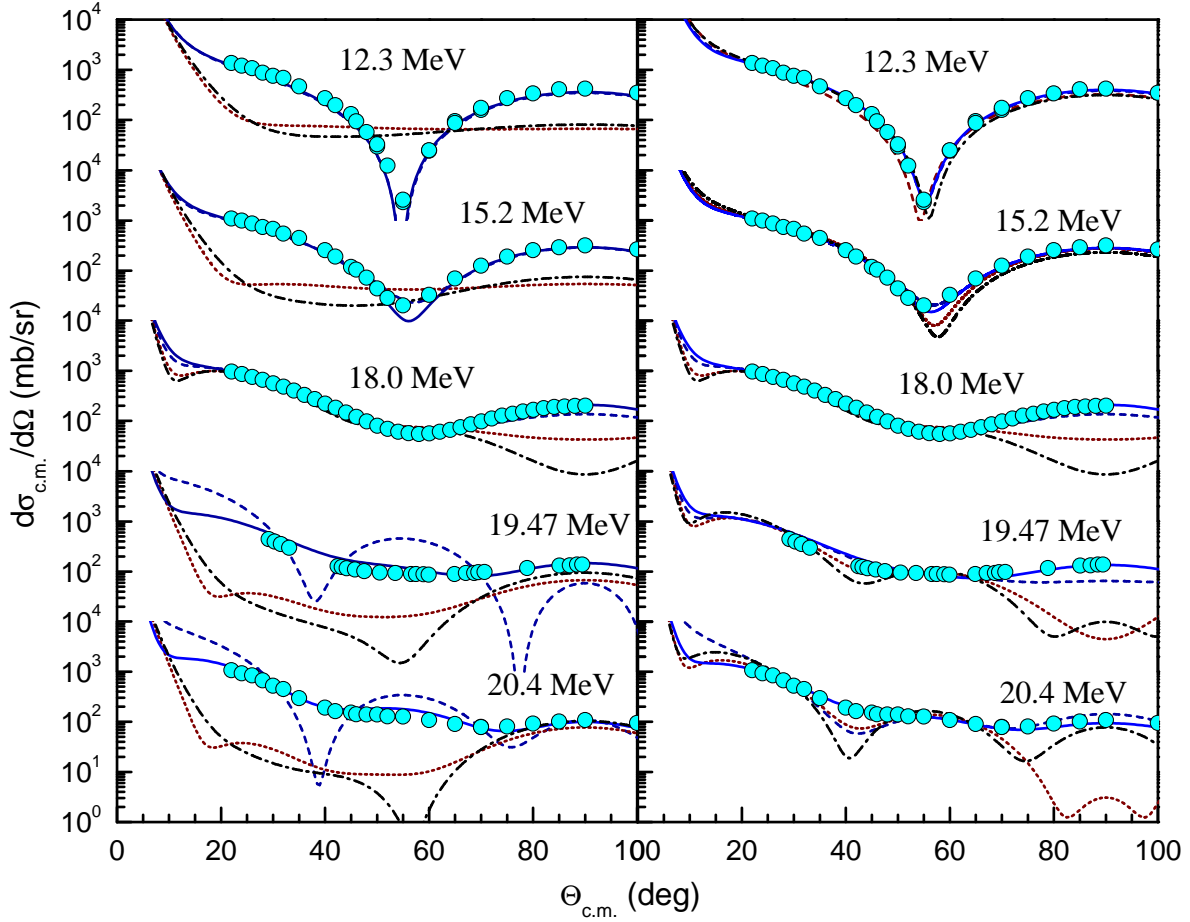}\vspace{-4cm}
\caption{The OM description of the elastic \aa data at energies 
of $E_\alpha=12.3$ to 20.4 MeV given by the \emph{unrenormalized} (left panel) and 
\emph{renormalized} (right panel) CDM3Y6 folded potentials obtained with different 
approximations for the density dependence. The notation of curves is the same as that 
in the left panel of Figs.~\ref{f5} and \ref{f6}.}  \label{f7} 
\end{figure}
\begin{figure}[bht]
 \vspace{3.5cm}\hspace{0cm}
\includegraphics[angle=0,scale=0.65]{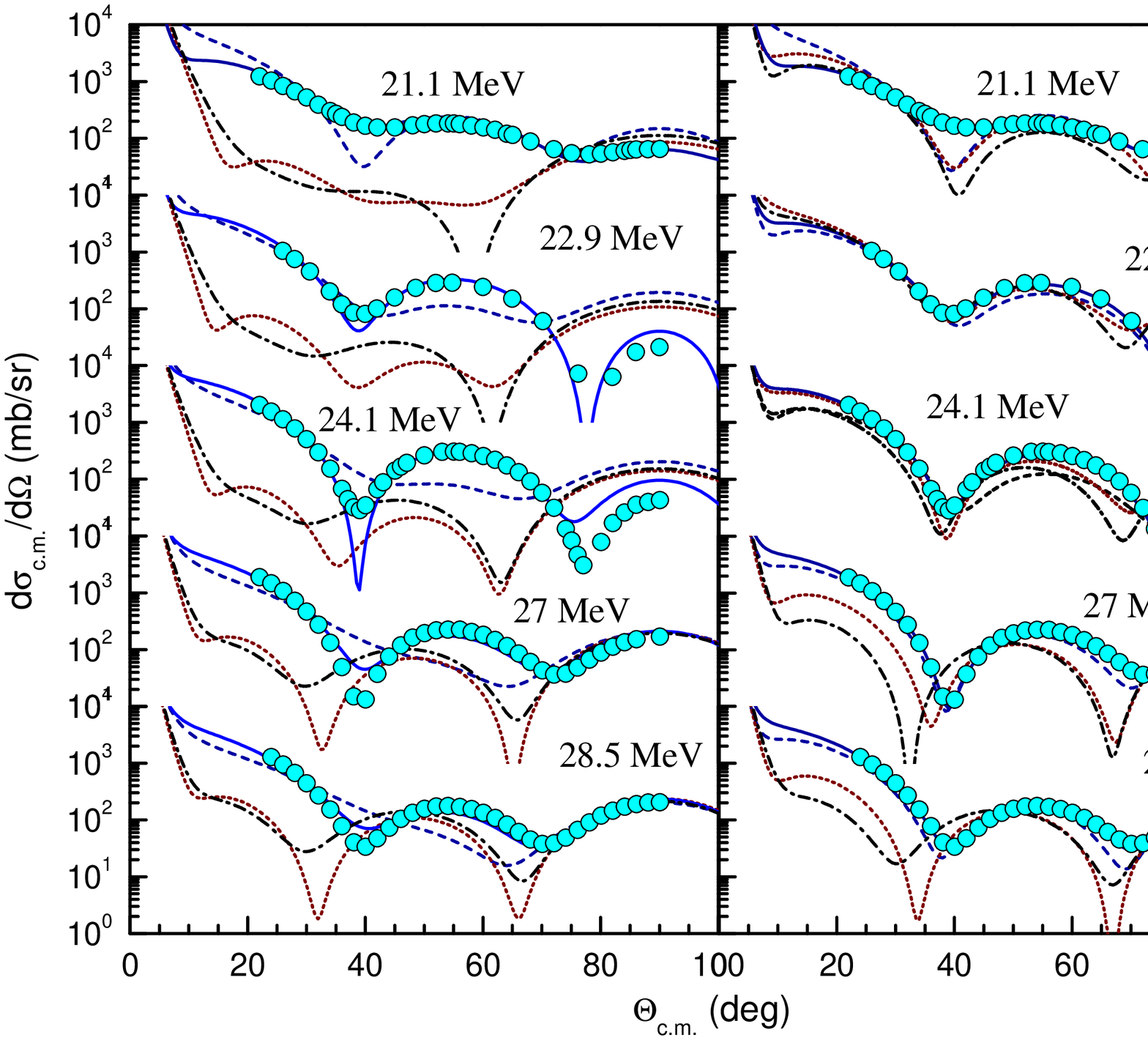}\vspace{-4cm}
\caption{The same as Fig.~\ref{f7} but for the energies $E_\alpha=21.1$ 
to 28.5 MeV.} \label{f8}
\end{figure}
The low-energy data at 8.87 MeV have a simple diffractive structure of one deep 
minimum that can be well described by all the folded potentials after some  
renormalization of the potential strength. However, one can see from 
Table~\ref{t1} that the renormalization coefficients found for the folded
potentials obtained with the FDA and AOD approximations are  
$N\approx 0.6$, much smaller than that found for the folded potential obtained 
with the DPD approximation $N\approx 0.96$. The 29.5 MeV data have a more 
complicated oscillation structure that can be reproduced only by the Gaussian 
potential and the CDM3Y6 potential obtained with the DPD approximation. 
Without the absorption, the considered elastic \aa scattering data are sensitive to 
the \aa potential down to quite small impact parameter. As an illustration, the OM  
description of the 8.87 and 29.5 MeV data given by the Gaussian potential 
of Buck {\em et al.}, with and without the contribution of the $L=0$ partial wave, 
are plotted in Fig.~\ref{f6k}. It is obvious that these data are quite sensitive to 
the partial wave $L=0$ that corresponds to the interaction distance of around 
1 fm between the two \al-particles (based on the semi-classical relation 
$L+1/2\approx kR$, where $k$ is the wave number). At such short distances,  
the folded potentials obtained with the FDA and AOD approximations are too shallow
and unable to account for the data at 29.5 MeV even after a $\chi^2$-fit of the 
renormalization factor $N$. Given a best-fit $N$ coefficient quite close to unity 
that gives a very good OM description of both data sets, the CDM3Y6 potential 
obtained with the DPD approximation is undoubtedly the best prediction of the 
\aa potential by the present folding model. The Gaussian potential of Buck {\em et al.} 
also gives a good description of the data at these energies, with the best-fit $N$ 
coefficient close to unity (see Table~\ref{t1}). A very important characteristic of the 
OP is the volume integral per interacting nucleon pair $J_V$ that has 
been often used to identify a given potential family \cite{Bra97}. It can be seen from 
Table~\ref{t1} that the CDM3Y6 potential obtained with the DPD approximation for the 
density dependence turned out to have the best-fit $J_V$ value very close to that 
of the Gaussian potential of Buck {\em et al.}. These two potentials seem to belong 
to the same deep potential family (with $-J_V\approx 400$ MeV~fm$^3$) found from the 
global systematics of the light ion elastic scattering (see, e.g., Fig.~6.7 of Ref.~\cite{Bra97}). 
The FDA and AOD approximations for the density dependence result on a too shallow 
folded potential whose $J_V$ value is significantly lower than that of the folded potential 
obtained with the DPD approximation. The fact that the folded potentials obtained 
with the FDA and AOD approximations could describe the 8.87 MeV data well after
being renormalized by $N\approx 0.6$ shows that the renormalized potentials  
have slipped into a very shallow potential family that gives a "phase-equivalent" 
description of elastic scattering at this energy. The shallow families of the
\aa potential were found long ago from the phase shift analysis of the low energy data
\cite{Ali}, but these shallow potentials are empirical and cannot be associated with 
the results of a microscopic model like RGM or DFM.  

\begin{figure}[bht]
 \vspace{3.5cm}\hspace{0cm}
\includegraphics[angle=0,scale=0.65]{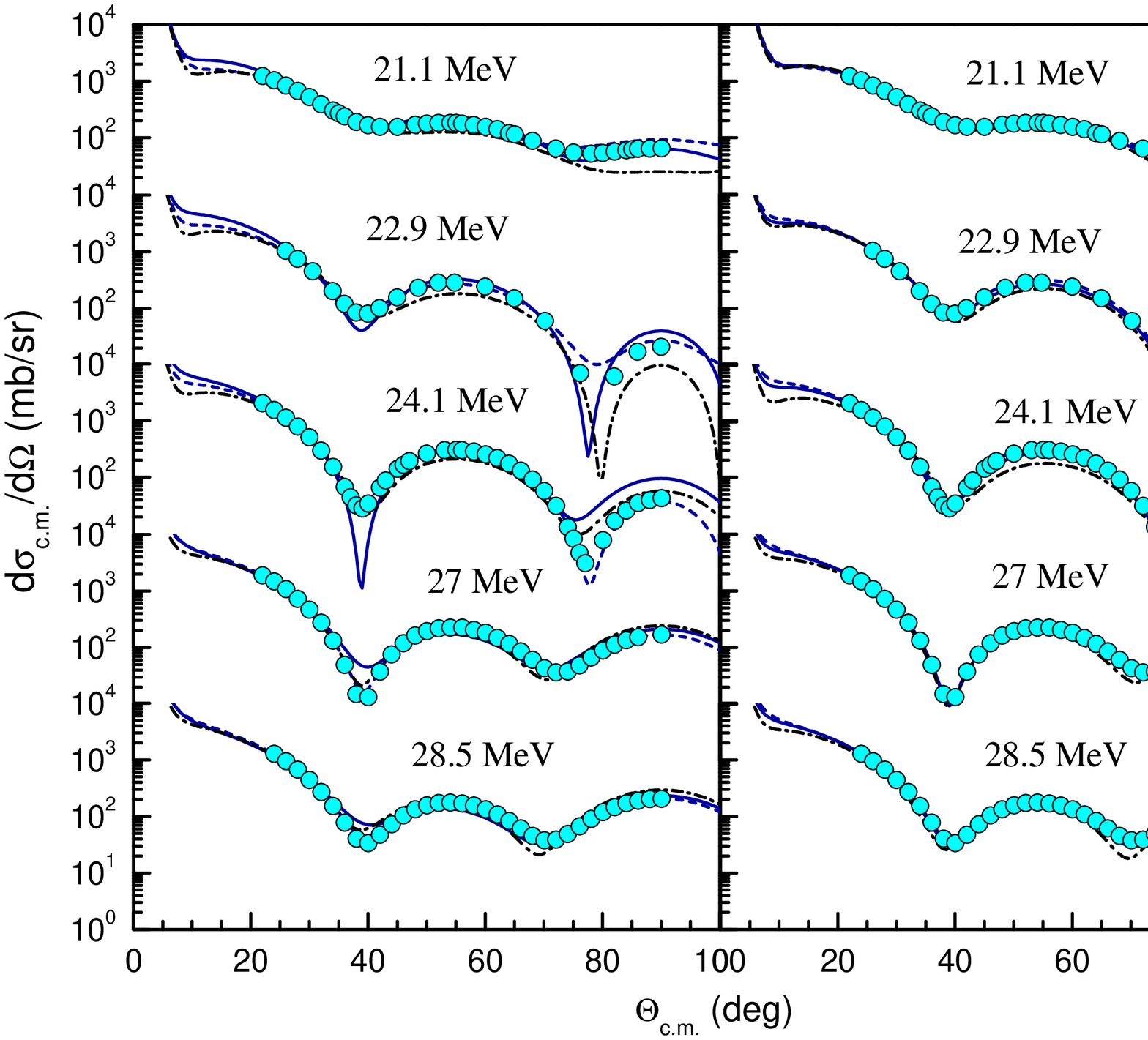}\vspace{-4cm}
\caption{The OM description of the elastic \aa data at energies 
of $E_\alpha=21.1$ to 28.5 MeV given by the \emph{unrenormalized} (left panel) and 
\emph{renormalized} (right panel) \aa potentials. The results given by the CDM3Y6 
and BDM3Y1 folded potentials obtained with the DPD approximation are shown as 
solid and dash-dotted curves, respectively. The dash curves are the results given 
by the Gaussian potential of Buck {\em et al.} \cite{Buck}.} \label{f9}
\end{figure}
\begin{figure}[bht]
 \vspace{3.5cm}\hspace{0cm}
\includegraphics[angle=0,scale=0.65]{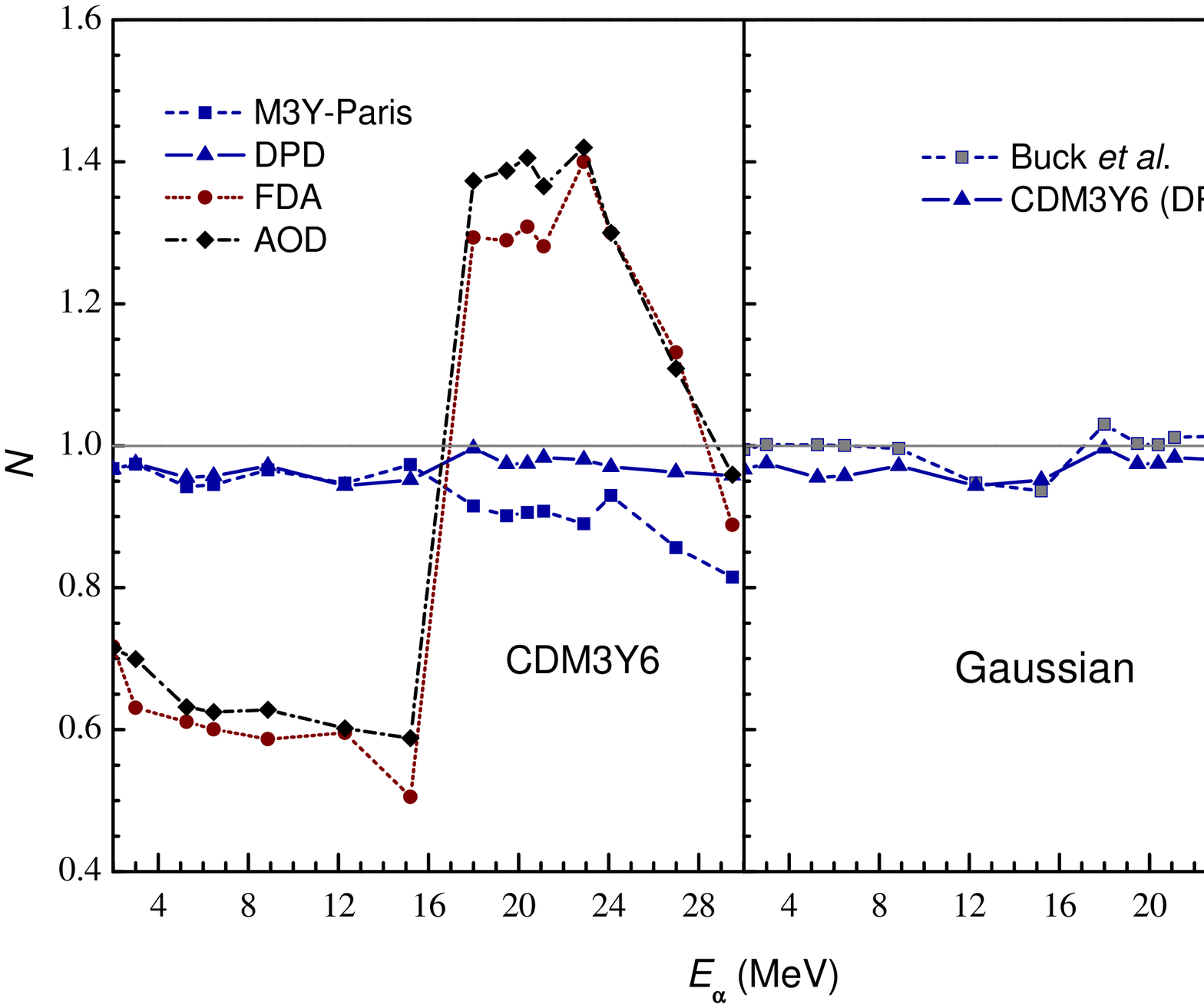}\vspace{-5cm}
\caption{The best-fit renormalization coefficient $N$ of the CDM3Y6 
folded potentials obtained with different approximations for the density dependence 
and the Gaussian potential of Buck {\em et al.} \cite{Buck} found by the OM analysis 
of the elastic \aa data at energies of $E_\alpha=3$ to 29.5 MeV.} \label{f10}
\end{figure}
For the simplicity, we discuss further only the DFM results obtained with 
the Gaussian density (\ref{e1a}). The effects caused by different approximations 
for the density dependence to the DFM description of the elastic \aa data at 8.87 
and 29.5 MeV were also confirmed in the OM analysis of the data at energies 3.0 MeV 
$\leqslant E_\alpha\leqslant 29.5$ MeV, using the folded \aa potentials. The elastic 
\aa data plotted in Figs.~\ref{f7} and \ref{f8} show clearly the evolution of the 
elastic \aa angular distribution (with the increasing energy) from that having a single 
diffractive minimum to a more complicated oscillating structure with two pronounced 
diffractive minima. The (unrenormalized) folded potentials obtained with the FDA and AOD
approximations fail completely to account for the data at all energies. After some 
renormalization of the potential strength, the single-minimum shape of the elastic cross 
section measured at low energies could be reproduced by all the folded potentials 
(see right panel of Fig.~\ref{f7}), but those given by the FDA and AOD approximations 
remain unable to describe the data at higher energies where the diffractive pattern 
is more complicated (see Fig.~\ref{f8}). The description of the data by the potential 
obtained with the density independent M3Y interaction also becomes worse with the 
increasing energy, and the only folded potential that gives consistently a good 
description of the data at all energies is that obtained with the density dependent 
M3Y interaction and DPD approximation for the density dependence. Given a 
renormalization coefficient $N\approx 0.96\sim 0.98$, the CDM3Y6 potential obtained 
with the DPD approximation delivers a very good description of all the considered data 
(see right panels of Fig.~\ref{f7} and \ref{f8}). From the two density-dependent 
versions of the folded potential, the BDM3Y1 potential gives a slightly poorer fit 
to the data compared to that given by the CDM3Y6 potential (see Fig.~\ref{f9}). 
The best-fit OM results given by these two folded potentials are compared 
with those given by the Gaussian potential of Buck {\em et al.} \cite{Buck} 
in Fig.~\ref{f9}. With the Gaussian parameters adjusted to reproduce the \aa 
resonance at the energy around 92 keV and the elastic phase shift \cite{Buck}, 
the Gaussian potential gives an excellent OM description of the elastic \aa data 
under study. One can see also from Fig.~\ref{f10} that the best-fit renormalization 
factor $N$ of the Gaussian potential of Buck {\em et al.} is close to unity in most 
cases, except for the two energies of 12.3 and 15.2 MeV where $N$ is falling below 0.95. 
Thus, we conclude that the potential of Buck {\em et al.} \cite{Buck} is a very realistic 
analytical form available for the local \aa potential at low energies. 
       
\begin{figure}[bht]
 \vspace{3.5cm}\hspace{0cm}
\includegraphics[angle=0,scale=0.65]{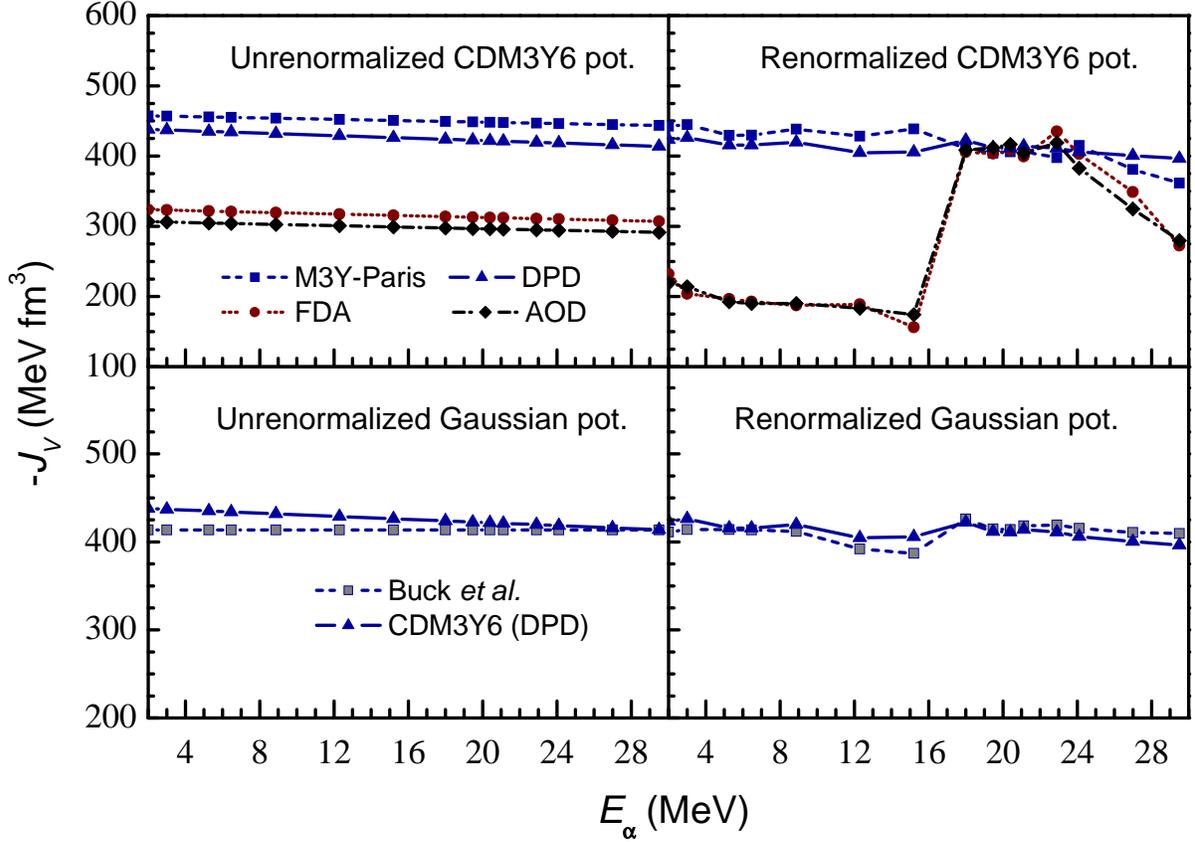}\vspace{-5cm}
\caption{The volume integral per interacting nucleon pair $J_V$ of the 
\emph{unrenormalized} (left panel) and \emph{renormalized} (right panel) CDM3Y6 
folded potentials obtained with different approximations for the density dependence 
and the Gaussian potential of Buck {\em et al.} \cite{Buck}. The $J_V$ values in the 
right panel are given by the \aa potentials renormalized by the corresponding $N$ 
coefficients shown in Fig.~\ref{f10}.}
 \label{f11} \end{figure}
\begin{figure}[bht]
 \vspace{3.5cm}\hspace{0cm}
\includegraphics[angle=0,scale=0.65]{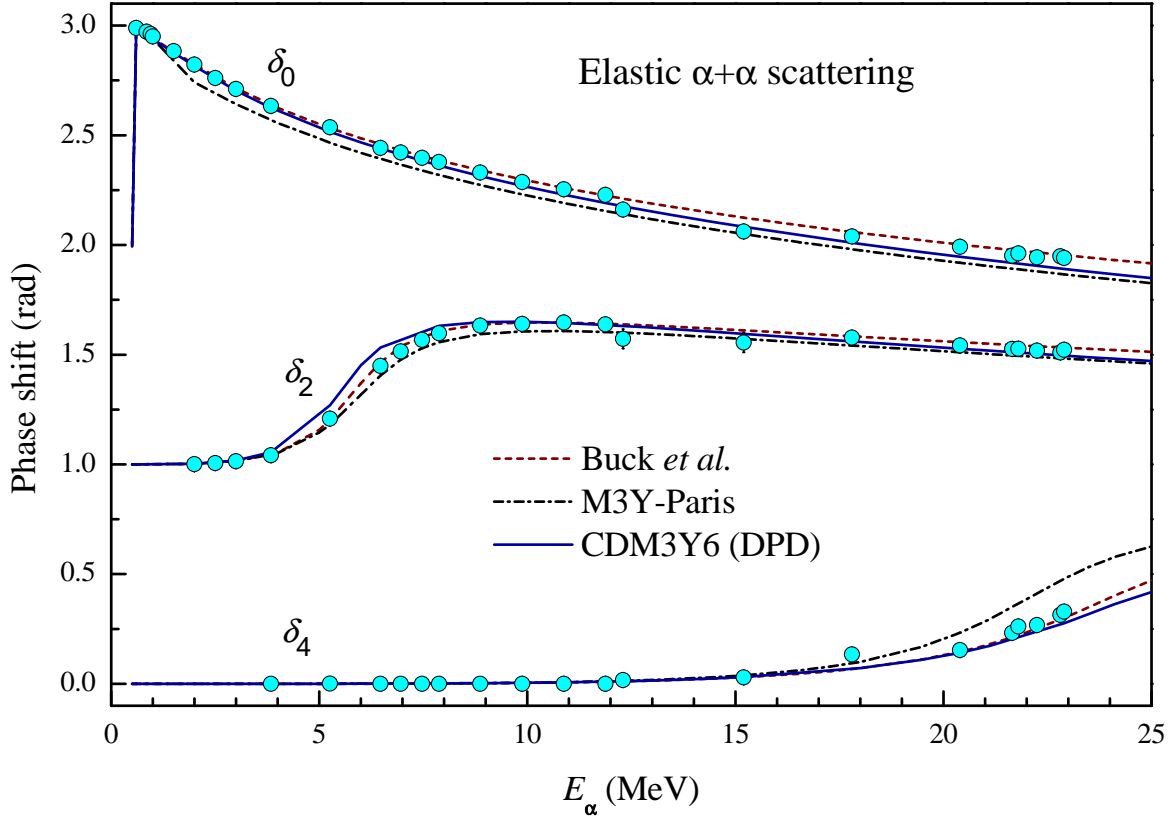}\vspace{-5cm}
\caption{The experimental elastic \aa phase shift $\delta_0,\ \delta_2$ 
and $\delta_4$ at low energies \cite{Afzal} in comparison with the OM results given by the 
(unrenormalized)  Gaussian potential of Buck {\em et al.} \cite{Buck} and the folded 
potentials obtained with the density independent M3Y-Paris interaction and 
density dependent CDM3Y6 interaction (using the DPD approximation). 
The M3Y-Paris and CDM3Y6 folded potentials are renormalized by $N=0.94$ and 0.97, 
respectively.}   \label{f12}
\end{figure}
The best-fit renormalization factor $N$ of the CDM3Y6 (DPD) potential turned out to be 
also close to unity as shown in Fig.~\ref{f10}. The volume integrals per interacting 
nucleon pair $J_V$ of different folded potential are compared with those given by the 
Gaussian potential (\ref{e13}) in Fig.~\ref{f11}, and one can see that the CDM3Y6 (DPD)
potential and the Gaussian potential belong indeed to the same potential family. 
Moreover, these two versions of the \aa potential also give the best OM description 
of the elastic \aa scattering data under study. Thus, we conclude that the DPD 
approximation is a much more accurate treatment of the density dependence (\ref{e3}) 
of the M3Y interaction compared to the FDA and AOD procedures. Although the DPD 
approximation leads to a very strong depletion of the overlap density in the center 
of the \aa system as shown in Fig.~\ref{f2}, a total neglect of the density dependence 
worsens somewhat the OM fit to the data, especially, the \aa scattering data measured 
at energies around 20 MeV and higher (see left panels of Figs.~\ref{f7} and \ref{f8}).

The present study has shown a vital role of the elastic scattering data measured over 
the whole observable angular range in testing different theoretical models for the 
\aa potential. The elastic phase shifts deduced from the phase-shift analysis of these 
experimental cross sections are widely referred to in the literature as the 
{\em experimental} phase shifts. These quantities are, however, less sensitive to the 
detailed shape and strength of the potential compared to the measured elastic
angular distribution. As illustration, the experimental elastic phase shift $\delta_0,\ 
\delta_2$ and $\delta_4$ at low energies \cite{Afzal} are compared with the OM results 
given by the Gaussian potential of Buck {\em et al.} \cite{Buck} and the 
folded potentials in Fig.~\ref{f12}. One can see that the Gaussian potential gives a 
perfect fit to the experimental phase shifts at all energies, although the detailed OM analysis 
has shown (Fig.~\ref{f10}) that it needs to be renormalized by a factor below 0.95 to fit 
the elastic data measured at energies of  12.3 and 15.2 MeV. The CDM3Y6 potential 
(obtained with the DPD approximation) renormalized by an average $N$ factor of 0.97 
gives about the same good fit to the elastic phase shift, while the density independent 
M3Y-Paris potential (renormalized by an average $N$ factor of 0.94) fails to describe 
the phase shift data at energies above 20 MeV, in agreement with the OM results shown 
in Figs.~\ref{f7} and \ref{f8}.        

\section{Summary}
The OM analysis of the elastic \aa scattering at energies below the reaction 
threshold of 34.7 MeV has been done using the folded potentials obtained with 
the CDM3Y6  and BDM3Y1 versions of the density dependent M3Y interaction 
\cite{Kho95,Kho97}. Different approximations for the density dependence (\ref{e3})
of the chosen interactions have been tested in the OM calculation and the results were   
compared with those given by the Gaussian potential of Buck {\em et al.} \cite{Buck} 
that is based on the microscopic RGM results. The elastic \aa data, in terms of the elastic 
scattering  cross section measured accurately over the whole observable angular range, 
were shown to be a very efficient test ground for different theoretical models 
of the \aa potential. 

Our consistent folding model description of the elastic \aa data under study has shown 
that the DPD approximation, based on the dynamic Pauli  distortion of the two 
overlapping \al-densities in the momentum space, gives the best folding model prediction 
of the \aa potential at low energies. The CDM3Y6 folded potential obtained with the 
DPD approximation turned out to be very close in strength and shape to the Gaussian 
potential of Buck {\em et al.} that was constructed to reproduce the experimental elastic 
\aa phase shifts and the RGM results for $^8$Be resonance  \cite{Buck}. 
These two potentials were shown to belong to the same potential family in terms 
of the potential depth and volume integral. They also give equally good description 
of the elastic \aa data under study. The present OM analysis of the elastic \aa scattering 
has also confirmed that the Gaussian potential of Buck {\em et al.} is a reliable analytical 
expression for the local \aa optical potential at low energies. This result validates, therefore, 
the use of this Gaussian potential as the \al-\al\ interaction in the \al-cluster folding 
calculation \cite{Ela01,Kar06}.  

Although the DPD treatment of the density dependence (\ref{e3}) of the M3Y 
interaction is just a local approximation based on the Pauli blocking of the overlap 
of the two \al-densities in a dinuclear matter picture in the momentum space, the 
results of the present study show clearly that the density dependence of the effective 
NN interaction should be strongly distorted at the small distances between the two 
\al-particles. This results seems to explain a long standing problem of the DFM. 
Namely, the failure of the DFM in a consistent description of both the \aa and 
\aN optical potentials at low energies using the same realistic density dependent 
NN interaction is due to a breakdown of the FDA by the Pauli blocking.  
The DPD approximation for the density dependence of the M3Y interaction can also 
be applied to study different nucleus-nucleus systems within the same dynamic DFM 
approach, and it is expected to improve the performance of the DFM at low energies, 
near the Coulomb barrier. We plan to carry out this topical research in the near future.

\section*{Acknowledgments}
The authors thank Yoshiko Kanada-En'yo for her helpful communication. 
The present research has been supported by the National Foundation for Science \&
Technology Development (NAFOSTED project No. 103.04-2011.21) and 
by the LIA program of the Ministry of Science and Technology (MOST).

\newpage
\appendix*
\subsection*{Appendix: Antisymmetrized overlap density of the \aa system}
We consider two $\alpha$ particles, whose centres of mass are separated by a distance 
$R$, in the Brink's microscopic cluster model for $^8$Be \cite{Hor10}. 
The antisymmetrized total wave function of the \aa system in this model is given 
explicitly as
\begin{equation}
 \Psi(R)=n_0(R) {\mathcal A}\{\psi_{\alpha_1}\times \psi_{\alpha_2}\}=
 \frac{n_0(R)}{\sqrt{4!4!}} \det \{\varphi_1 \varphi_2 ... \varphi_8 \}, \label{eq1}
\end{equation}
where the single-nucleon state is determined as the s$_{1/2}$ harmonic oscillator
wave function
\begin{eqnarray}
\varphi_i &=& g \left({\bm r}_i+\frac{\bm R}{2},b\right)\chi_i \xi_i 
 \quad {\rm if} \quad i=1,...,4 \nonumber\\
\varphi_i &=& g \left({\bm r}_i-\frac{\bm R}{2},b\right)\chi_i \xi_i 
\quad {\rm if} \quad i=5,...,8 \nonumber\\ g ({\bm r},b)&=& \frac{1}{\pi^{3/4}b^{3/2}} 
\exp\left(-\frac{{\bm r}^2}{2b^2}\right). \label{eq3}
\end{eqnarray}
The h.o. parameter $b=1.1932$ fm \cite{Kho01}, $\mathcal{A}$ is the antisymmetrizing 
operator, $\chi_i$ and $\xi_i$ are the spin and isospin wave functions of the $i$-th nucleon, 
and $n_0(R)$ is the normalization constant determined from the condition
\begin{equation}
\langle \Psi(R)\mid \Psi(R)\rangle = n_0(R)^2 \frac{8!}{4!4!} \langle \Psi_8 \mid  
{\mathcal A} \Psi_8 \rangle=n_0(R)^2 \frac{8!}{4!4!} \det B=1. \label{eq4} 
\end{equation}
Here $\Psi_8 = \varphi_1...\varphi_8,\  {\mathcal A}\Psi_8=
 \det \{\varphi_1...\varphi_8\}$, and the matrix elements of $B$ are 
\begin{eqnarray}
 B_{ij}=\langle\varphi_i|\varphi_j \rangle &=&  1 \qquad\qquad \qquad {\rm if}\ j=i, \nonumber\\
   &=& \exp\left(-\frac{R^2}{4b^2}\right) \  \ {\rm if}\ j=i\pm 4, \nonumber\\
   &=& 0 \qquad\qquad\qquad   {\rm otherwise}. \label{eq5}
\end{eqnarray}
The antisymmetrized overlap density of the \aa system is determined as
\begin{eqnarray}
\rho({\bm r},R)&=&\langle \Psi(R)\mid \sum_{i=1}^{8} \delta({\bm r}-{\bm r}_i)\mid
\Psi(R)\rangle \nonumber \\
 &=& n_0(R)^2 \frac{8!}{4!4!} \langle \Psi_8 \mid \sum_{i=1}^{8} \delta({\bm r}-{\bm r}_i) 
 \mid {\mathcal A} \Psi_8 \rangle \nonumber \\
&=& \sum_{i=1}^8 \sum_{j=1}^8 \langle\varphi_i | \delta({\bm r}-{\bm r}_i) | 
 \varphi_j \rangle (B^{-1})_{ji}, \label{eq6}
\end{eqnarray}
where the matrix elements of the $\delta$-function are
\begin{eqnarray}
\langle \varphi_i|\delta({\bm r}-{\bm r}_i)|\varphi_j \rangle   
 &=& \frac{1}{\pi^{3/2} b^3} \exp \left[ -\frac{1}{b^2} \left({\bm r}\pm 
\frac{\bm R}{2}\right)^2 \right] \ {\rm if}\ j=i, \nonumber\\ 
&=& \frac{1}{\pi^{3/2} b^3} \exp \left[-\frac{1}{b^2} \left(r^2 +\frac{R^2}{4}
\right) \right] \ \ \ {\rm if}\ j=i\pm 4, \nonumber \\
   &=& 0 {\hskip 5.5cm}   {\rm otherwise}. \label{eq7}
\end{eqnarray}
In the diagonal matrix elements, the sign is (+) when $i=j\leqslant 4$ and (-) 
when $i=j\geqslant 5$. Using the explicit  expression for $(B^{-1})_{ji}$ derived 
directly from Eq.~(\ref{eq5}), we obtain the \aa overlap density (\ref{eq7}) in the 
following compact form
\begin{eqnarray}
\rho({\bm r},R)&=&\frac{4}{\pi^{3/2}b^3[1-h(R,b)]}\Big\{
\exp\left[-\frac{1}{2b^2}\left(\bm{r}-\frac{\bm R}{2}\right)^2\right]
-2\exp\left(-\frac{r^2}{b^2}\right)h(R,b) \nonumber \\
&+&\exp\left[-\frac{1}{2b^2}\left(\bm{r}+\frac{\bm R}{2}\right)^2\right]
 \Big\},\ \mbox{with}\ h(R,b)=\exp\left(-\frac{R^2}{2b^2}\right).  \label{eq8}
\end{eqnarray}
   
Like the h.o model for the \al-density discussed in 
Sec.~\ref{sec1}, this model for the \aa overlap density can also be corrected for 
the center-of-mass motion using the prescription of Ref.~\cite{Tas58}, and the h.o. 
range of the Gaussians in Eq.~(\ref{eq3}) will be modified accordingly. However, 
all the results would be the same if we keep using the same empirical h.o. parameter 
$b=1.1932$ fm in both cases without c.m. motion correction. 
 
\bigskip

\end{document}